\documentclass[nofootinbib,superscriptaddress,
               amsmath,amssymb,floatfix]{revtex4}
\usepackage{pstricks,pst-node,amsmath,amssymb}

\usepackage [dvips]{graphicx}
\usepackage{amsmath}
\usepackage{epsfig}
\usepackage{amssymb}

\newcommand{\vol}{N}

\newcommand{\C}{\mathcal C}

\newcommand{\KS}{\text{KS}}

\newcommand{\dd}{\text{d}}

\newcommand{\ee}{\text{e}}

\begin{document}

\title{First-order dynamical phase transition in models of glasses:
  an approach based on ensembles of histories}

\author{Juan P. Garrahan}

\affiliation{School of Physics and Astronomy, University of
Nottingham, Nottingham, NG7 2RD, UK}

\author{Robert L. Jack}

\affiliation{Department of Chemistry, University of California,
Berkeley, CA 94720-1460}
\affiliation{Department of Physics, University of Bath,
Bath BA2 7AY, UK}

\author{Vivien Lecomte} \affiliation{Laboratoire Mati\`ere et
  Syst\`emes Complexes (CNRS UMR 7057), Universit\'e Paris Diderot, 10
  rue Alice Domon et L\'eonie Duquet, 75205 Paris cedex 13, France}
\affiliation{D\'epartement de Physique de la Mati\`ere Condens\'ee,
  Universit\'e de Gen\`eve, 24 quai Ernest-Ansermet, 1211 Gen\`eve,
  Switzerland}
			
\author{Estelle Pitard}

\affiliation{Laboratoire des Collo\"{\i}des, Verres et
Nanomat\'{e}riaux (CNRS UMR 5587), Universit\'e de Montpellier II,
place Eug\`ene Bataillon, 34095 Montpellier cedex 5, France}

\author{Kristina van Duijvendijk}

\affiliation{Laboratoire Mati\`ere et Syst\`emes Complexes (CNRS UMR
  7057), Universit\'e Paris Diderot, 10 rue Alice Domon et L\'eonie
  Duquet, 75205 Paris cedex 13, France}

\author{Fr\'ed\'eric van Wijland}

\affiliation{Laboratoire Mati\`ere et Syst\`emes Complexes (CNRS UMR
  7057), Universit\'e Paris Diderot, 10 rue Alice Domon et L\'eonie
  Duquet, 75205 Paris cedex 13, France}

\begin{abstract}
We investigate the dynamics of kinetically constrained models of glass
formers by analysing the statistics of trajectories of the dynamics, or histories, using
large deviation function methods.  We show that, in general, these models exhibit a first-order dynamical transition between active and inactive dynamical phases.  We argue that the dynamical
heterogeneities displayed by these systems are a manifestation of dynamical first-order phase coexistence. 
In particular, we calculate dynamical large deviation functions, both analytically and numerically, for the Fredrickson-Andersen model, the East model, and constrained lattice gas models.  We also show how large deviation functions can be obtained from a Landau-like theory for dynamical fluctuations.  We discuss possibilities for similar dynamical phase-coexistence behaviour in other systems with heterogeneous dynamics.
\end{abstract}

\maketitle

\section{Introduction}

In this paper we describe in detail a theoretical method for the study of the dynamics of glassy systems~\cite{glasses1,glasses2,glasses3,glasses4}.  This approach is in essence a statistical mechanics of the trajectories of the dynamics, or {\em histories}, as it is based on the study of large deviation functions~\cite{Touchette}---which can be thought of as generalized free-energies---of dynamic observables.  In particular, we use the tools of Ruelle's thermodynamic formalism~\cite{Ruelle,GaspardBook}, as applied to continuous time Markov chains \cite{FormaThermo}, to study kinetically constrained models (KCMs) of glass formers \cite{Ritort-Sollich}.  In a recent letter~\cite{KCMdyntransition} we showed using these methods that the dynamics of KCMs takes
place on a first-order coexistence line between active and inactive
dynamical phases, in accordance with previous suggestions~\cite{Merolle-Jack}.  Here we expand significantly on
Ref.\ \cite{KCMdyntransition}, demonstrating in detail the existence of the first-order
dynamical phase transition, and discussing the Landau-like
approach~\cite{FormaThermoLandauFreeEn} that we use to characterise the dynamical phases, and the transition between them.   The dynamical transition we find in KCMs \cite{KCMdyntransition} is related neither to a thermodynamic transition, nor to a finite temperature (or finite density) dynamical singularity.  Our results, therefore, point towards a perspective \cite{Merolle-Jack} on glasses which is distinct from other approaches, such as the random first-order transition theory \cite{Mezard-Parisi,Franz-Parisi,Xia-Wolynes,mosaic}, frustration-limited domains \cite{Tarjus}, or  mode-coupling theory \cite{MCT}. 

The paper is organized as follows: in Section~\ref{sec:sstates}
we introduce our dynamical tools and the ensemble of
histories in which the dynamical phase transition takes place.  In Section~\ref{sec:models} we describe 
the models that we will consider.  We show the existence of
a dynamical phase transition in Section~\ref{sec:results}
comparing different models and establishing 
minimal conditions that are sufficient to ensure a dynamical
transition.  In Section~\ref{sec:traj} we discuss
the ensemble of histories in detail, considering
statistical properties of the active and inactive phases,
and a dynamical analogue of phase separation.  We summarise
our results in Section~\ref{sec:concl}, and consider
some open questions.

\section{Dynamical tools: the $s$-ensemble}
\label{sec:sstates}

\subsection{Motivations}

In this article, we are concerned with fluctuations in dynamical
observables such as the amount of dynamical activity in a glassy
system, integrated over a long time $t$ and over a large (but finite)
system.  To investigate these fluctuations, we consider 
statistical properties of the histories followed by the system.
Ensembles of histories are central to
the thermodynamic formalism developed
by Ruelle and coworkers~\cite{Ruelle} (see~\cite{GaspardBook} for a
comprehensive review).   While thermodynamics is concerned with 
probability distributions over configurations of a large system, 
we will apply the thermodynamic formalism to probability distributions
over histories.
We begin by discussing the physical content of the observables
that we will consider.

In the Boltzmann-Gibbs theory, the macroscopic features of large
systems are characterised by determining the
statistical properties (the mean value and fluctuations) of extensive
observables, such as the energy or the number of particles.
In a \emph{microcanonical} approach, one considers the properties
of a system with fixed total energy $E$.  They are obtained from 
the counting factor
\begin{equation} \label{eq:OmegaE}
  { \Omega(E,\vol)} \ = \ 
  \left|
    \begin{aligned}
      &\text{number of configurations}\\
      &\text{with energy $E$}\\
    \end{aligned}
  \right.
\end{equation}
where $\vol$ represents the size (the volume) of the system. 
In the large size limit ($\vol\to\infty$), we define the entropy density
$s(e)=\lim_{\vol\to\infty} \frac 1\vol \ln \Omega(e\vol,\vol)$, which
represents the relative weight of configurations with energy density~$e$.

In a dynamical context, we consider histories of
the system between an initial time $\tau=0$ and a final time $\tau=t$.
Instead of considering the statistics of the energy $E$, we will
consider an observable $A$, that is extensive in the observation
time $t$.  
The dynamical analog of $\Omega(E,\vol)$ is the probability
distribution of this observable
\begin{equation} \label{eq:Omega_dyn}
\Omega_\mathrm{dyn}(A,t) = \left|
    \begin{aligned}
      &\text{fraction of histories with a given value of the}\\
      &\text{time-extensive observable $A$}\\
    \end{aligned}
  \right.
\end{equation}
On a mathematical level, the choice of the observable $A$ is 
somewhat arbitrary, although application of 
the thermodynamic formalism requires that the quantity
$
\frac 1t \log \Omega_\mathrm{dyn}(at,t)
$
should have a finite limit for large times $t$. 
Subject to this constraint, the choice of the order parameter
$A$ is informed by physical insight:
we should use an observable that reveals the essential
physical processes at work in the system.  For example, in
non-equilibrium systems in contact with two reservoirs of particles, 
we might define $A$ as the total particle current:
the number of particles transferred from
one reservoir to the other between times $0$ and $t$ (see,
for example, Refs~\cite{Giardina,clones}).
In the context of glassy phenomena, we consider observables 
that measure the ``activity'' or the ``complexity'' of the 
history~\cite{FormaThermo,Merolle-Jack,KCMdyntransition}.

Returning to the Boltzmann-Gibbs approach,
it is useful to define the \emph{canonical} ensemble through
the partition function
\begin{equation}\label{eq:Z_boltz}
  {  Z(\beta,\vol)} \ = \
  \sum_{E}  \Omega(E,\vol)\,\text{e}^{-\beta {E}} 
 \end{equation}
which characterises a system at a given temperature $\beta^{-1}$.
Within this framework, phase transitions can be identified from 
singularities in the intensive free energy,
$f(\beta)=-\lim_{\vol\to\infty} \frac {1}{\beta \vol} \ln Z(\beta,\vol)$.
The dynamical analog of this thermodynamic partition
sum is 
\begin{equation}
\label{eq:Z_dyn}
  {  Z_A(s,t)} \ = \
  \sum_{A}  \Omega_{\text{dyn}}(A,t)\, \text{e}^{-s {A}}
\end{equation}
where we introduced an intensive field $s$ conjugate to $A$.
This field will play a role similar to the inverse temperature
$\beta$. The dynamical partition function $Z_A(s,t)$ is the
central object of Ruelle's thermodynamic formalism.  

We have focused on the correspondence between the
thermodynamic limit of large system size ($\vol\to\infty$) 
and the long time limit ($t\to\infty$) in Ruelle's formalism.  
In the following, we will consider systems for which the
large time limit is to be taken at fixed system size: in some cases,
we will then take a second limit of large system size $\vol$.  
If we consider systems with no thermodynamic phase transitions,
then no singular behaviour arises on taking the
limit of large $\vol$ at fixed $t$.  In this case, 
we expect the limits of large $\vol$ and large $t$ to commute, but 
this is is clearly not the case in general.

\subsection{Systems with Markov dynamics: statistics over histories}

\subsubsection{Continuous time Markov evolution}

We now give more precise definitions of the quantities discussed
so far, by reviewing the construction of the
ensemble of histories for stochastic systems.
We focus on continuous-time Markov
dynamics~(in this section, we follow \cite{FormaThermo}). 
The system is defined by a finite set
of configurations $\{\C\}$.  Its dynamical evolution is 
defined by the rates $W(\C\to\C')$ for transitions from configuration $\C$ to
configuration $\C'$.  Thus, the
probability $P(\C,t)$ of being in configuration $\C$ at time $t$
evolves according to a master equation: 
\begin{equation} \label{eqn:cont_time_Markov_eq}
  \partial_t P(\C,t) = -r(\C)P(\C,t)+
    \sum_{\C'}  W(\C'\to\C)P(\C',t) 
\end{equation}
where
\begin{equation}\label{opevolmatel2}
 r(\C)=\sum_{\C'}W(\C\to \C')
\end{equation}
represents the rate of escape from $\C$. 
Equ.\eqref{eqn:cont_time_Markov_eq} is sufficiently
general to describe kinetically constrained models, spin
facilitated models, or lattice gases, with $\C$ 
representing the configuration of the whole lattice in each case.

\begin{figure}[t]
\begin{center}
\epsfig{file=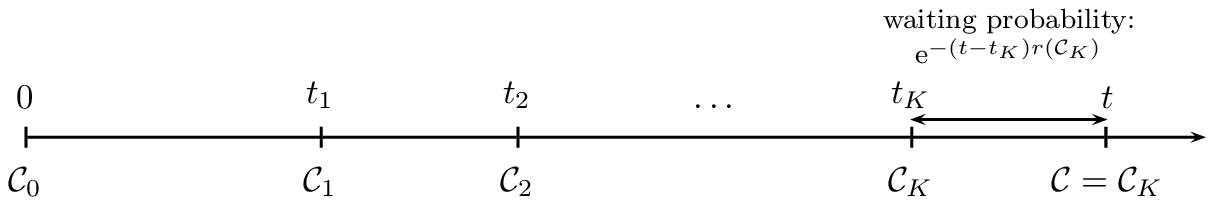}
\end{center}
\caption{A history of duration $t$ is defined by
a sequence of configurations $\C_0\to\ldots\to\C_K$ and
  a sequence of jump times $t_1,\ldots,t_K$.
  Between $t_{k}$ and $t_{k+1}$, the system
  stays in  configuration $\C_k$.  
}
\label{fig:time_sequence}
\end{figure}

Starting from a configuration $\C_0$ at initial time $t=0$, the
system will experience a fluctuating number 
of changes of configuration (``jumps'') between $0$ and $t$. 
We shall refer to the number of jumps as the ``activity'' 
and denote it by $K$.
A history (or trajectory) consists of a sequence $\C_0\to\ldots\to\C_K$ of
visited configurations, and  a sequence of times
$t_1,\ldots,t_K$ at which the jumps occur (Fig.~\ref{fig:time_sequence}).
We stress that for a fixed observation time $t$, the number of jumps is
a fluctuating quantity: it depends on the particular history followed by the
system between $0$ and $t$.  We refer to histories with many hops (large $K$)
as `active' histories and those with few hops (small $K$) as `inactive'. 

We use the notation $\langle\mathcal{O}\rangle$ for an average 
of the observable $\mathcal{O}$, over histories of the system.  
We consider observables that depend on the entire history of
the system, through the configurations visited and the time spent in each:
that is, $\mathcal{O}=\mathcal{O}(\C_0\dots\C_K,t_1\dots t_K)$.
In general, we have
\begin{equation} \label{eq:hist_measure}
\langle\mathcal{O}\rangle 
 = 
\sum_K \sum_{\C_0\dots \C_K} \int\!\dd t_1\dots\dd t_K\, 
p_0(\C_0)
\left[\prod_{k=1}^K W(\C_{k-1}\to\C_k) \right] 
  \exp\left[ -\sum_{k=1}^K r(\C_k)(t_{k+1}-t_k) \right] 
\mathcal{O}(\C_0\dots\C_K,t_1\dots t_K)
\end{equation}
where the limits on the time integrals are $t_1>0$, $t_{k}<t_{k+1}$, 
and $t_K<t_{K+1}\equiv t$;
we use $p_0(\C_0)$ to denote the probability
distribution of the initial configuration $\C_0$.
We use a compact notation for averages of this form:
\begin{equation} 
\langle\mathcal{O}\rangle = \sum_\mathrm{hist} \mathrm{Prob}[\mathrm{hist}]
\mathcal{O}[\mathrm{hist}].
\end{equation}
where
$\mathrm{Prob}[\mathrm{hist}]$ 
plays the role of a probability density in the
space of histories.

\subsubsection{Time-extensive observables}
\label{sec:obsA_B}

Having defined our system and its histories, we now turn
to the choice of the time-extensive observable $A$.  A simple choice
of this observable will be the activity $K$.  
Each time
the system changes configuration $\C\to\C'$ the activity $K$
is incremented: $K\to K+1$.
More generally, we can consider an observable $A$ that is incremented
at each jump, with the increment $\alpha(\C,\C')$ depending
on the configurations before and after the jump.
That is, for a given history with $K$ changes of configurations
\begin{equation} \label{eq:obsAsum}
  A[\text{hist}]=\sum_{k=0}^{K-1}\alpha(\C_{k},\C_{k+1})
\end{equation}
Again, we note that if $\alpha(\C,\C')=1$ then $A$ is the activity $K$.

To construct the dynamical partition sum, we start with
a `microcanonical' approach, classifying trajectories by their values 
of $A$.  We generalise the probability $P(\C,t)$, defining 
$P(\C,A,t)$ as the probability of being in configuration 
$\C$ at time $t$,
having measured a value $A$ of the time-extensive variable between $0$
and $t$.  Its evolution in time is given by
the master equation
\begin{equation}\label{eq:evolutionA}
  \partial_t P(\C,A,t) = \sum_{\C'} 
  W(\C'\rightarrow\C) P\big(\C',A-\alpha(\C',\C),t\big)\ - r(\C) P(\C,A,t)
\end{equation}

Thus, the probability of measuring a value $A$ for
the observable $A$ in a history of length $t$ is
\begin{equation}\Omega_\mathrm{dyn}(A,t)\equiv \sum_\C P(\C,A,t)
\end{equation} 
which we identify as the quantity introduced in~\eqref{eq:Omega_dyn}.  

\subsubsection{Canonical description: evolution in the $s$-ensemble}
\label{sec:s_ensemble_obsA_B}

We have defined the distribution $\Omega_\mathrm{dyn}(A,t)$  that 
is the analog of the microcanonical
counting factor $\Omega(E,\vol)$.  We now introduce the analog for
the canonical (Boltzmann-Gibbs) ensemble, parameterized by a field
$s$.  This involves a modification to the statistical weight
of each history: 
\begin{equation} \label{eq:def_s_ens}
\mathrm{Prob}[\mathrm{hist}] \to 
\mathrm{Prob}[\mathrm{hist}] \ee^{-sA[\mathrm{hist}]}
\end{equation}
Thus, in the `$s$-ensemble', averages of
observables $\mathcal{O}$ are given by
\begin{equation} \label{eq:def_ave_s_ensemble}
  \langle \mathcal O\rangle_{s} = \frac{1}{Z_A(s,t)} 
\sum_\mathrm{hist}
\mathcal{O}[\mathrm{hist}]
\mathrm{Prob}[\mathrm{hist}] \ee^{-sA[\mathrm{hist}]}
=
  \frac{\langle \mathcal O\, \ee^{-sA} \rangle }
  {\langle \ee^{-sA} \rangle }
\end{equation}
where
\begin{equation} \label{eq:def_Z_A_Laplace}
  Z_A(s,t) = 
\sum_\mathrm{hist} 
       \mathrm{Prob}[\mathrm{hist}] \ee^{-sA[\mathrm{hist}]}
           = \langle \ee^{-sA} \rangle
\end{equation}
is the dynamical partition function, introduced in~\eqref{eq:Z_dyn}.
(The subscript $A$ of $Z_A$ serves as a reminder that
the field $s$ is conjugate to $A$.)

Averages in the ensemble with $s=0$ correspond to the steady state averages
of $\mathcal O$. {\it A priori}, this is the only physically accessible 
ensemble.  Positive or negative values of $s$ favor histories with
non-typical values of $A$.  
%
For our purposes, working in the $s$-ensemble is
simpler than considering ensembles with fixed values of $A$.
We take the Laplace transform of $P(\C,A,t)$ with respect to $A$:
\begin{equation} \label{eq:def_P_hat_Cst}
  \hat{P}_A(\C,s,t)=\sum_A\ee^{-s A}P(\C,A,t)
\end{equation}
From~\eqref{eq:evolutionA},
the equation of
motion for $\hat{P}_A(\C,s,t)$ is
\begin{equation}  \label{eqn:evol_hatP}
  \partial_t \hat P_{A}(\C,s,t) = \sum_{\C'} 
  \ee^{-s\alpha(\C',\C)} W(\C'\to\C) \hat P_{A}(\C',s,t)\ - r(\C) \hat P_{A}(\C,s,t),
\end{equation} 
or, in an operator notation,
$
  \partial_t \hat {P}_A=\mathbb{W}_A\hat{P}_A
$\,,
where $\mathbb{W}_A$ operates in the space 
of configurations $\{\C\}$.  Its matrix elements are
\begin{equation}
  \label{eqn:operatorA}
  \big(\mathbb{W}_A\big)_{\C,\C'}=W(\C'\to\C)\ee^{-s\alpha(\C',\C)}-r(\C)\delta_{\C,\C'} .
\end{equation}
Some properties of the operator $\mathbb{W}$ are discussed in 
appendix~\ref{app:eigenvecW_A}: Equ.~\eqref{eq:P_long_t} states that
$\hat P_{A}(\C,s,t)$ behaves in the large time limit as
$
   \hat{P}_{A}(\C,s,t) \sim R_0(\C,s) \ee^{t\psi_{A}(s)} 
$
where $\psi_{A}(s)$ is the largest eigenvalue of $\mathbb{W}_A$ 
and $R_0(\C,s)$ is the associated eigenvector. 
Thus, for large times,
\begin{equation} \label{eq:Z_A_conv}
Z_A(s,t)
=\sum_\C \hat{P}_A(\C,s,t)
\sim\ee^{t\psi_{A}(s)},\end{equation}
and we will refer to $\psi_{A}(s)$ as (the negative
of) the \emph{dynamical free energy} per
unit time.  Summing Eq.~\eqref{eq:def_P_hat_Cst} over $\C$, 
probability conservation implies $Z_A(0,t)=1$,
so that $\psi(0)=0$ for all stochastic systems.

\subsubsection{Large deviation functions}
\label{sec:large_dev}

In the Boltzmann-Gibbs theory, entropy and free energy are related through
a Legendre transform (as can be seen from~\eqref{eq:Z_boltz} or \cite{Kuboetal}) which
provides a link between microcanonical and canonical ensembles. 
We have already defined the function $\psi_K(s)$, which is the
dynamical analog of the 
free energy density $f(\beta)$.
The dynamical analog of the entropy density $s(e)$ is
\begin{equation}
\pi(a) = \lim_{t\to\infty} \frac 1t \log \Omega_\mathrm{dyn}(at,t)
\end{equation}
which determines the large-$t$ scaling of the probability of observing a 
value $at$ for the observable $A$.

For large times, the sum in~\eqref{eq:Z_dyn} is dominated
by the maximum of $\Omega_\mathrm{dyn}(A,t)$, so
that $\pi(a)$ and $\psi_A(s)$ are 
are related through a Legendre transform:
\begin{align}
   \psi_{A}(s)&=\max_{a} \big(\pi(a)-sa\big)
  \label{eq:s_vs_a} 
\end{align}
If the function $\pi(a)$ satisfies $\pi''(a)\leq0$, it
can be obtained from the inverse transform
\begin{align}
   \pi(a)&=\min_{s} \big(\psi_{A}(s)+sa\big)
  \label{eq:a_vs_s}
\end{align}

Physically, the quantity $\pi(a)$ describes the large fluctuations of $A$. 
It is maximal at the most probable value of $a$, which is the mean value
of $A/t$, in the limit of large time $t$. 
Gaussian fluctuations of $A/t$ are described by the quadratic
approximation of $\pi(a)$ around its maximum.
Expanding $\pi(a)$ beyond quadratic order gives information about
non-Gaussian fluctuations of $A/t$, which are
referred to as \emph{large deviations} \cite{Touchette}.  Alternatively,
one may characterise these fluctuations through
$\psi_{A}(s)$, since the cumulants of $A$ are obtained from the derivatives
of $\psi_A(s)$ through
$
  \lim_{t\to\infty } \frac 1t \langle A^{p}\rangle_\mathrm{c} = 
  (-1)^{p}\left.\frac{\dd^{p}\psi_{A}(s)}{\dd s^{p}}\right|_{s=0}
$, where, as usual, $\langle A^{p}\rangle_\mathrm{c}$ is the $p$-th cumulant of $A$.

\subsubsection{Time-extensive observables varying continuously in time}
\label{sec:obs_B}

In addition to time-extensive order parameters of the form
given in~\eqref{eq:obsAsum}, we also consider those of the form 
\begin{equation} \label{eq:obsBsum}
B[\mathrm{hist}] 
        = \sum_{k=0}^{K} (t_{k+1}-t_k) b(\C_K) 
= \int_0^t dt'\: b(\C(t')), 
\end{equation}
where we introduced a configuration-dependent observable $b(\C)$.
In the sum over $k$, we define $t_0=0$ and $t_{K+1}=t$ so that
the time spent in configuration $\C_k$ is simply $t_{k+1}-t_k$.  
In the integral representation, we have represented the
trajectory by a function
$\C(t')$ which takes the value $\C_k$ for $t_k<t'<t_{k+1}$.
The time-integrated energy of the system is an observable
of the form $B$, in which case $b(\C)$ is simply the energy of 
configuration $\C$.
Then, defining $P(\C,B,t)$ by analogy with $P(\C,A,t)$, we have
\begin{equation}\label{eq:evolutionB}
  \partial_t P(\C,B,t) = \sum_{\C'} 
  W(\C'\rightarrow\C) P\big(\C',B,t\big)\ - r(\C) P(\C,B,t) - b(\C) \frac{\partial}{\partial B} P(\C,B,t)
\end{equation}
We define an $s$-ensemble associated with the observable
$B$ through 
\begin{equation}
\mathrm{Prob}[\mathrm{hist}] \to 
\mathrm{Prob}[\mathrm{hist}] \ee^{-sB[\mathrm{hist}]}
\end{equation}
Then, repeating the analysis of Section~\ref{sec:s_ensemble_obsA_B},
the analog of $\psi_A(s)$ is  
$\phi_B(s)=\lim_{t\to\infty} \frac 1t \ln \langle\ee^{-sB}\rangle$.
This quantity is equal to the maximal 
eigenvalue of an operator $\mathbb W_B$, whose elements are
\begin{equation} \label{eqn:operatorB}
  \left(\mathbb W_{B}\right)_{\C,\C'} = 
   W(\C'\to\C) - \big[ r(\C)+s b(\C) \big] \delta_{\C,\C'}
\end{equation}
In the following, we concentrate our study on time-extensive variables
of type $A$.  Some connections
between $s$-ensembles parameterized by observables of 
types $A$ and $B$ discussed in Appendix~\ref{app:AB}

\subsubsection{Variational approach for $\psi_A(s)$}
\label{sec:variational_approach}

The models considered in this work have dynamics which obey
detailed balance 
with respect to an equilibrium distribution $P_{\text{eq}}(\C)$:
that is, $P_{\text{eq}}(\C) W(\C\to\C')= P_{\text{eq}}(\C') W(\C'\to\C)$.
This allows us to derive a variational bound on the dynamical
free energy $\psi_K(s)$.
To achieve this, we symmetrise the
 evolution operator $\mathbb W_K$, defining
$\tilde{\mathbb  W}_K$
through the similarity transformation
$(\tilde{\mathbb  W}_K)_{\C,\C'} = P_{\text{eq}}^{-1/2}(\C) (\mathbb  W_K)_{\C,\C'} 
P_{\text{eq}}^{1/2}(\C')$. 
Hence, 
\begin{equation} \label{eq:symmOp}
 \big(  \tilde{\mathbb  W}_K\big)_{\C'\C} = 
 \ee^{-s}\big[W(\C\to\C')W(\C'\to\C)\big]^{\frac 12}-r(\C)\delta_{\C \C'}=\big(  \tilde{\mathbb  W}_K\big)_{\C\C'}
\end{equation}
Since $\tilde{\mathbb W}_K$ and ${\mathbb W}_K$ are related by
a similarity transformation, their eigenspectra are
identical.  We therefore use
a variational principle (valid for any symmetric operator) to determine 
their common maximal eigenvalue:
\begin{equation} \label{eq:max_P}
  \psi_{K}(s)
= \max_{\{V(\C)\}} \frac{ \sum_{\C,\C'}
V(\C) (\tilde {\mathbb W}_{K})_{\C,\C'} V(\C')}
{\sum_\C V(\C)^2} 
= \max_{|V\rangle} 
\frac{\langle V|\tilde {\mathbb W}_{K}|V\rangle}{\langle V|V\rangle}
\end{equation}
At $s=0$, the maximum is achieved for 
$V(\C)=P_\mathrm{eq}(\C)^{1/2}$, and $\psi(0)=0$, as required.

Interestingly, the quantity to be maximised in~\eqref{eq:max_P}
has a physical interpretation.  For any history
of the system, the fraction of time spent
in each configuration $\C$ defines a quantity
known as the experimental measure.  As we discuss
in appendix~\ref{app:DonskerVaradhan}, Donsker-Varadhan theory relates 
the  probability
of observing a particular experimental measure to
an expectation value of the form 
$\langle V|\tilde{\mathbb W}_K|V\rangle$.  
In Section~\ref{sec:traj}, we will use these results to investigate
fluctuations in the $s$-ensemble.

\section{Models and order parameters}
\label{sec:models}

\subsection{Kinetically constrained models: FA, East, TLG and KA models}

Kinetically constrained models \cite{Fredrickson-Andersen,Jackle,Kob-Andersen,Jackle-tlg,Kurchan-Peliti,Sollich-Evans,Einax_superactivated,Garrahan-Chandler,Aldous_Diaconis,YJ04,Toninelli-Biroli-Fisher,Pan-tlg,Geissler-Reichman} are simple lattice models of glasses which can account for a large range of dynamical phenomena associated to the glass transition.  This includes: super-Arrhenius temperature dependence of timescales, non-exponential relaxation, spatially heterogeneous dynamics, transport decoupling, and aging and memory effects. The thermodynamic properties of KCMs are very simple, and their
non-trivial features arise from dynamical rules which
forbid or favor some transitions, while maintaining detailed balance
with respect to a trivial equilibrium distribution over configurations.
For a review on KCMs see \cite{Ritort-Sollich}.

We first consider models with
binary spins $n_i=0,1$ where $i=1,\dots,N$ are the sites of a lattice.  
In spin-facilitated models, sites with $n_i=1$
represent excitations, which promote
local activity.  The models evolve by single spin-flips, 
which occur with rates
\begin{equation}
W(n_i \to 1-n_i) = C_i(\{n_j\}) \frac{\ee^{\beta(n_i-1)}}{1+\ee^{-\beta}}
\end{equation}
where $\beta$ is the inverse temperature, and
the kinetic constraint enters through the function
$C_i(\{n_j\})$, which is a function of the neighbors 
$n_j$ of $i$, but does not itself depend on $n_i$.  In
this case, it is simple to verify that the model
obeys detailed balance with respect to the equilibrium
distribution
\begin{equation} \label{eq:Peq_hard}
P_\mathrm{eq}(\{n_i\}) = \prod_i \frac{\ee^{-\beta n_i}}{1+\ee^{-\beta}}
\end{equation}
and that the excitation density is
$c\equiv\langle n_i \rangle=(1+e^\beta)^{-1}$.

In the one-spin facilitated Fredrickson-Andersen (FA)
model~\cite{Fredrickson-Andersen,Ritort-Sollich}, $C_i=1$ if any of
the nearest neighbors $j$ of $i$ are in the excited state,
$n_j=1$; otherwise $C_i=0$.  We also 
consider the three-dimensional variant of 
the East model~\cite{Jackle,Sollich-Evans,Berthier-Garrahan} 
in which $C_i=1$
for site $i=(x,y,z)$ if at least one of the sites
$(x-1,y,z)$, $(x,y-1,z)$ or $(x,y,z-1)$ is in the excited state;
otherwise $C_i=0$.

In addition, we consider lattice gas 
models~\cite{Kob-Andersen,Jackle-tlg,Ritort-Sollich},
in which particles move from site to site, with
at most one particle per site.
Sites which are occupied have $n_i=1$, and unoccupied
sites have $n_i=0$.
Particles move between sites $i$ and $j$ with rate
$C_{ij}(\{n_k\})$ so that
the model has a conserved density $\rho=\vol^{-1}\sum_i n_i$.
The rate $C_{ij}(\{n_k\})$ is
non-zero only for nearest neighbor sites $i$ and $j$, and   
it is independent of $n_i$ and $n_j$.
Thus, equilibrium state has a trivial distribution: 
all configurations with density $\rho$ have equal probability.
As an example of such a model, 
we consider the two-vacancy facilitated triangular
lattice gas, or (2)-TLG
\cite{Jackle-tlg}, which is defined on a triangular lattice, 
with a constraint $C_{ij}$ which is equal to unity if the two
common nearest neighbors of sites $i$ and $j$ are vacant, and
zero otherwise.  Similarly, the (2,2) variant of the
Kob-Andersen (KA) lattice model~\cite{Kob-Andersen}
 is defined on a square lattice, with $C_{ij}=1$
if at least one neighbour $k\neq j$ of site $i$ has $n_k=0$
and at least one neighbour $k'\neq i$ of site $j$ has $n_{k'}=0$.
Otherwise $C_{ij}=0$. 

\subsection{Reducibility of KCMs and sums over histories}
\label{sec:reducibility}

The construction of the $s$-ensemble in Section~\ref{sec:sstates}
assumed
that the system of interest has a single steady state to
which it converges in the long-time limit.  For finite-sized
stochastic systems, this convergence is ensured as long as
the dynamics are irreducible~\cite{Ritort-Sollich}:
that is, it must be possible for
every configuration of the system to be reached from every
other configuration.  For KCMs, this is not the case in general.
For example, in the FA model, there are no transitions either
into or out of the configuration with no excited sites ($n_i=0$ 
for all $i$).
For the models considered here, these states
are usually considered to be irrelevant because they have a 
contribution to the Gibbs measure that vanishes exponentially
in the thermodynamic limit, at all temperatures $T>0$.

However, when considering large deviations, these states may 
become relevant.  In order to ensure convergence to a single
steady state, we define our unbiased measure over histories
as in~\eqref{eq:hist_measure}, with a distribution
of initial conditions $p_0(\C_0)$ that is non-zero only
for configurations in the largest irreducible partition
of the dynamics.  That is, we do not allow the system to
occupy configurations that cannot be reached from representative
configurations taken from the relevant Gibbs ensemble.  For
the FA model and East models, this simply means that the
system may not occupy the configuration which has no excited
sites.  Practically, this means (for example)
that the maximum in the variational expression~\eqref{eq:max_P}
should be taken with the constraint that
$V(\C)$ is finite only for configurations
in the largest irreducible partition.

Instead of restricting initial conditions
in this way, one could instead consider large deviations
in a subsystem of size $\vol$ that is embedded in a larger 
system of size $\vol'\gg \vol$: this was the approach taken
in~\cite{Merolle-Jack}.  As usual, we expect these approaches
to be equivalent in the limit of large system size $\vol$.

\subsection{Kinetically constrained models : bosonic and mean-field variants}
\label{sec:def_bos_FA}

It is convenient to
define a `bosonic' variant of the one-spin facilitated FA model \cite{Whitelam,Jack-Mayer}, 
in which $n_i$ may be any integer greater than or equal to zero.  We take 
\begin{equation}
W(n_i\to n_i+1) = C_i(\{n_j\}) \ee^{-\beta},
  \qquad
W(n_i\to n_i-1) = C_i(\{n_j\}) n_i
\end{equation}
where $C_i$ is again independent of $n_i$ so that 
\begin{equation} \label{eq:Peq_boson}
P_\mathrm{eq}(\{n_i\})= \prod_i \frac{c^{n_i} }
{n_i!}
\ee^{-c}
\end{equation}
where $c\equiv\langle n_i \rangle = \ee^{-\beta}$.
For the bosonic FA model in finite dimension, we take
$C_i(\{n_j\})=\sum_{\langle j\rangle}n_j$, where the sum is over
the nearest neighbours $j$ of site $i$.

For this bosonic model, it is convenient to use
the Doi-Peliti 
representation~\cite{Doi-Peliti}.  We define bosonic
operators $a_i$, $a_i^\dag$ and $\hat n_i=a^\dag_i a_i$,
with $[a_i,a^\dag_j]=\delta_{ij}$ 
and a vacuum state $|0\rangle$ for which $a_i|0\rangle=0$ for all $i$.
The Doi-Peliti representation of the operator $\mathbb W$ is 
defined by $(\mathbb W)_{\C\C'}=\langle 0|\left[\prod_i 
\frac{a_i^{n_i}}{n_i!}\right]
\mathbb W \prod_i (a_i^\dag)^{n_i'}|0\rangle$  where the configurations
$\C$ and $\C'$ have occupations $\{n_i\}$ and $\{n_i'\}$ respectively.
In the $s$-ensemble, we are interested in the operator $\mathbb W_A$
defined in~\eqref{eqn:operatorA}.  In the case where the observable
$A$ is the activity $K$, we have 
\begin{equation}
\mathbb{W}_K^\mathrm{(FA)} = \sum_{\langle ij\rangle} \left[
 \ee^{-s} ( c a_i^\dag + a_i ) \hat n_j + 
 \ee^{-s} ( c a_j^\dag + a_j ) \hat n_i  - 
 2(\hat n_i \hat n_j + c) 
\right]
\end{equation}
where the sum runs over (distinct) pairs of nearest neighbours.

In addition, it is useful to consider a mean-field variant
of the FA model, in which the facilitation function of site
$i$ depends
symmetrically on the state of all sites.  That is,  
$W(n_i\to n_i+1)=N^{-1}\sum_j n_j e^{-\beta}$, and
$W(n_i+1\to n_i)=N^{-1}\sum_j n_j n_i$,
which satisfy detailed balance with respect 
to~\eqref{eq:Peq_boson}.
In the Doi-Peliti representation,
the master operator is simply
\begin{equation}
\mathbb{W}_K^\mathrm{(FA,mf)} = 
(2N)^{-1} \sum_{ij} \left[ 
\ee^{-s} a_i^\dag ( c a_j^\dag + a_j ) a_i  +
\ee^{-s} a_j^\dag ( c a_i^\dag + a_i ) a_j  
- 2 ( a_i^\dag a_j^\dag a_j a_i + c)  \right]
\end{equation}
Due to the symmetry between sites, the
properties of the model can be obtained from
a single co-ordinate: the total
number of excitations $n_\mathrm{tot}=\sum_i n_i$, whose
equilibrium distribution
is Poissonian with mean $cN$.  In this co-ordinate,
the master-like equation~\eqref{eqn:evol_hatP} has
a closed form, and
the matrix elements of the relevant operator are simply
\begin{equation}
(\mathbb{W}_K^\mathrm{(FA,mf)})_{n_\mathrm{tot}',n_\mathrm{tot}} = 
  c n_\mathrm{tot}(\ee^{-s}\delta_{n_\mathrm{tot}+1,n_\mathrm{tot}'}
 - \delta_{n_\mathrm{tot},n_\mathrm{tot}'}) 
  + \frac{n_\mathrm{tot}}{N}(n_\mathrm{tot}-1)(\ee^{-s}\delta_{n_\mathrm{tot}-1,n_\mathrm{tot}'}
 - \delta_{n_\mathrm{tot},n_\mathrm{tot}'})
\end{equation}

\subsection{The A-model and the AA model}

It will be useful to compare the FA model with two other
models,  which we call the A and AA models.  These names
are motivated by the
schematic representations of their fundamental processes, as
$A\leftrightarrow\emptyset$ and $A+A\leftrightarrow\emptyset$.
Here we have used an alternative notation to avoid confusion 
with the observable $A$ used to define the $s$-ensemble.

We define the A-model and its bosonic
variant by removing
the kinetic constraints from the FA model: that is, 
$C_i(\{n_j\})=1$, independent of the state
of the system.  In this model, excitations are 
created and destroyed singly, independent of site.
The A-model has the same equilibrium distribution
as the FA model, but its large deviations can be solved 
exactly. 

We also compare the FA model with a model in which particles
appear and annihilate (AA) in pairs~\cite{cardytauber}.  This so-called
AA model is related to a variant of
the FA model, through a  similarity transformation
that connects their master operators~\cite{Jack-Mayer}.
The AA model is defined for binary spins $n_i=0,1$.
In this model, the excitations move between adjacent sites with
rate $D$, and appear and annihilate in pairs with rates $k$ and $k'$ 
respectively.  Schematically, we write
\begin{equation} \label{eq:dynAA}
  0_i 1_j\stackrel{D}{\leftrightarrow}1_i0_j,\qquad
  1_i1_j\stackrel{k}\to0_i0_j,\qquad
  0_i0_j\stackrel{k'}\to 1_i1_j
\end{equation}
for neighbouring sites $i$ and $j$.
The equilibrium state 
of this model is of the form~\eqref{eq:Peq_hard}, with
$\ee^{-\beta}=\sqrt{k'/k}$.
It is also convenient to consider bosonic and mean-field
variants of this model, defined analogously to their
FA counterparts.  In the bosonic
AA model, we generalise to $n_i\geq0$, using rates
\begin{equation} \label{eq:bAA}
  (n_i, n_j)\stackrel{D n_i}{\rightarrow}(n_i-1, n_j+1),\qquad
  (n_i, n_j)\stackrel{k n_i n_j}\to (n_i-1, n_j-1),\qquad
  (n_i, n_j)\stackrel{k'}\to (n_i+1, n_j+1)
\end{equation}
which obey detailed balance with respect to~\eqref{eq:Peq_boson} with
$k'=k\ee^{-2\beta}$ as before.  
In the Doi-Peliti formalism, we have
\begin{equation}
\mathbb{W}_K^\mathrm{(AA)} = \sum_{\langle ij\rangle} 
  \ee^{-s} [ k' a_i^\dag a_j^\dag + k a_i a_j + D (a_i^\dag a_j +a_j^\dag a_i)]
- [k\hat n_i \hat n_j
 + ck' + D (\hat n_i + \hat n_j)]
\end{equation}

In the mean field variant of the AA model, 
diffusion occurs between all pairs
of sites ($i\neq j$), with rate $(D/N)$; pair
creation and annihilation processes occur with
rates $(k/N) n_i n_j$ and $(k'/N)$ for all pairs of sites $i\neq j$;
and we also allow for on-site pair creation and annihilation:
$n_i\to n_i\pm2$ with rates 
$(k'/N)$
and 
$(k/N) n_i (n_i-1)$. 
In the Doi-Peliti representation, the master operator is
\begin{equation}
\mathbb{W}_K^\mathrm{(AA,mf)} = \frac{1}{2N} \sum_{ij} \left[
  \ee^{-s} ( k' a_i^\dag a_j^\dag + k a_i a_j )  
- (k a_i^\dag a_j^\dag a_j a_i + ck')
 \right] 
 + \frac{D}{2N} \sum_{i\neq j} \left[ \ee^{-s} (a^\dag_i a_j + 
             a^\dag_j a_i) - (\hat n_i + \hat n_j) \right]
\end{equation}
For finite systems, the restriction to $i\neq j$ in the
diffusion term 
means that the master-like equation cannot be written
in terms of the single co-ordinate $n_\mathrm{tot}$, except at $s=0$.  
However, in the limit of large-$N$, this single co-ordinate
is sufficient, and the master-like operator for this co-ordinate
reduces to
\begin{equation}
(\mathbb{W}_K^\mathrm{(AA,mf)})_{n_\mathrm{tot}',n_\mathrm{tot}} = 
  k'N (z\delta_{n_\mathrm{tot}+2,n_\mathrm{tot}'}
 - \delta_{n_\mathrm{tot},n_\mathrm{tot}'}) 
  + k \frac{n_\mathrm{tot}}{N}(n_\mathrm{tot}-1)(z\delta_{n_\mathrm{tot}-2,n_\mathrm{tot}'}
 - \delta_{n_\mathrm{tot},n_\mathrm{tot}'})
 + D(z-1)n_\mathrm{tot}
\end{equation}
with $z=\ee^{-s}$.

\subsection{Relevant observables}

We now discuss the observables that we 
will use to define the $s$-ensemble, and those
that we will use to characterise trajectories within
that ensemble.

\subsubsection{The activity $K$ and the complexity $Q_+$}
\label{sec:activityK}

We have already defined the activity $K$, which counts the number
of changes of configuration in a dynamical trajectory.
In the context of dynamically heterogeneous systems such as 
glass-formers, the local activity can be used to distinguish mobile
and immobile regions of the system.  The large deviations of
the extensive activity $K$ are used to characterise trajectories
which are more or less mobile than average.

We note that $K$ is of the form given in~\eqref{eq:obsAsum} with
$\alpha(\C',\C)=1$, so the properties of the relevant $s$-ensemble
are encoded in the operator
\begin{equation}
  \big(\mathbb{W}_K\big)_{\C,\C'}=\ee^{-s}\,W(\C'\to\C)-r(\C)\delta_{\C,\C'} 
\end{equation}


Systems with dynamical heterogeneities are likely 
to present a wide distribution of very different histories. 
 One way of characterizing this diversity is provided by the
\emph{dynamical complexity} of the
histories~\cite{Glotzer,Merolle-Jack,gaspard1993}. In the context of dynamical
system theory, this quantity is called the Kolmogorov-Sinai entropy
\cite{Ruelle}. It provides one with the information
content of the history and is defined as the logarithm of the probability
of the history.
As discussed in~\cite{FormaThermo}, the appropriate generalisation
of this approach to systems with Markov dynamics is to consider
the entropy associated with the measure over sequences of 
configurations $\C_0\to\ldots\to\C_K$~\cite{FormaThermo}. 
This amounts to performing a coarse-graining in 
time: it means that the information associated with the 
time intervals between changes of configuration is ignored when
calculating the complexity.  The definition of the
dynamical complexity is 
\begin{equation}\label{defQ+}
  Q_+ = \sum_{k=0}^{K-1}\ln \frac{W(\C_k\to\C_{k+1})}{r(\C_k)} \ ,
\end{equation}
which is of the form given in~\eqref{eq:obsAsum}.
Thus, we define a dynamical partition sum
\begin{equation} \label{eqn:Z_as_generating_func}
  Z_{+}(s,t) =\langle \ee^{-s Q_+}\rangle.
\end{equation}
The corresponding dynamical free energy is
$
  \psi_+(s)
  = \lim_{t\to\infty} \frac{1}{t} \ln Z_{+}(s,t)  
$
which corresponds to the \emph{topological pressure} 
of dynamical system theory.  The analog of
the 
Kolmogorov-Sinai entropy $h_\KS$ is 
\begin{equation}
   h_\KS = - \lim_{t\to\infty} \frac{1}{t} 
             \left\langle Q_+ \right\rangle
         = \frac{\mathrm{d}}{\mathrm{d}s} \psi_+(s)
\end{equation}
which provides a measurement of the dynamical complexity of the
histories in the steady state.  In the examples of glass formers we
will study below, the dynamical ensembles given by $K$ and $Q_+$ are
qualitatively similar: we concentrate on the activity $K$
for simplicity.

\subsubsection{Fluctuation theorem in the $s$-ensemble}
\label{sec:fluctuation_theorem}

The Gallavotti--Cohen relation holds also in the quasi-stationary
state at fixed value of $K$ (or $s$), and therefore the
fluctuation--dissipation theorem holds there as well. In order to see
this, we parallel the reasoning presented by Lebowitz and Spohn in
\cite{lebowitzspohn}, and we construct the operator governing the
dynamics not only at fixed value of the activity $K$ but also at fixed
value of the entropy current
$Q_S=\sum_{n=0}^{K-1}\ln\frac{W(\C_n\to\C_{n+1})}{W(\C_{n+1}\to\C_n)}$,
which, in terms of the variables $s$ and $\lambda$ conjugate to the
activity $K$ and the entropy current $Q_S$ respectively, leads to the
following pseudo--evolution operator,
\begin{equation}
\Big(\mathbb{W}(s,\lambda)\Big)_{\C,\C'}
=\ee^{-s}W(\C'\to \C)^{1-\lambda}W(\C'\to \C)^\lambda-r(\C)\delta_{\C,\C'}
\end{equation}
whose property $\mathbb{W}(s,\lambda)^\dag=\mathbb{W}(s,1-\lambda)$
ensures that its largest eigenvalue $\psi$ verifies
$\psi(s,\lambda)=\psi(s,1-\lambda)$.  For system with particle
conservation, and subject to a field driving the system out of
equilibrium, we note that the entropy current $Q_S$ is directly
proportional to the total current of particles flowing through the
system~\cite{lebowitzspohn}.  In that case, the generalized symmetry
$\psi(s,\lambda)=\psi(s,1-\lambda)$ implies a fluctuation-dissipation
like relation in the $s$-ensemble.

\subsubsection{Order parameters within the $s$-ensemble}

As well as using the observables $K$ and $Q_+$ to define
$s$-ensembles through~\eqref{eq:def_s_ens}, we also characterise
the $s$-ensemble by using two other order parameters.
For spin-facilitated models, we consider the mean
excitation density:
\begin{equation}\label{def-moy-etat-s}
  \rho_K(s) \equiv \lim_{t\to\infty} \frac 1{\vol } \Big\langle \int_0^t \dd\tau\: \sum_i n_i(\tau) \Big\rangle_s
\end{equation}

For lattice gas models, the particle density is specified by the
initial conditions, so we require a
different order parameter. The average activity
is given by $\frac 1{\vol t}\langle K\rangle_s$.  One can also
consider the mean escape rate $r(\C)$ which depends only
on configurations of the system.  Again, we time-average this quantity
along the trajectories, and divide by the system size $\vol $, defining
\begin{equation}
  r_K(s) \equiv \lim_{t\to\infty} \frac 1{\vol t} \Big\langle \int_0^t \dd\tau\: r(\C(\tau)) \Big\rangle_s
\end{equation}

\section{Dynamical transitions in models of glass-formers}
\label{sec:results}

\subsection{Existence of a transition in KCMs: variational bounds}
\label{sec:phasecoexist}

It is clear from their equilibrium distributions $P_\mathrm{eq}(\C)$
that KCMs have no
phase transitions at any finite temperature.  That is,
their thermodynamic free energies are analytic functions of temperature
(or chemical potential).
However, we now show that in the limit of large 
time $t$ and large system size $\vol $, the dynamical free energy 
density $\vol ^{-1}\psi_K(s)$ has a singularity at $s=0$.  To
be precise, the dynamical free energy has a discontinuous first
derivative with respect to $s$, so we interpret this singularity
as a dynamical analog of a first-order phase transition.

The proof of such a transition is based on the escape rates
$r(\C)$ from the configurations of the model.  We establish
two bounds on $\psi(s)$.  Firstly, the number of
jumps $K$ is non-negative, so Eq.~\eqref{eq:def_Z_A_Laplace} 
implies that $Z_K(s,t)$ is a
non-increasing function of $s$.  Thus $\psi_K(s)$ 
is also non-increasing.  Further, $\psi_K(0)=0$, so we have
\begin{equation} \label{eq:bound_zero}
\psi_K(s)\leq 0, \qquad s\geq0
\end{equation}
Secondly, we can use the variational result~\eqref{eq:max_P} with
$V(\C_1)=1$ for just one configuration $\C_1$, and $V(\C)=0$
otherwise to establish
\begin{equation} \label{eq:bound_r}
\psi(s)\geq -\min_\C [ r(\C) ] 
\end{equation}
for all $s$.
%
For our purposes, the most important property of the kinetically
constrained models defined above is that they have 
\begin{equation} \label{eq:kcm_r}
\lim_{\vol \to\infty} \vol ^{-1} \min_\C r(\C)=0
\end{equation}  
This can be established by explicit construction.
In the FA and East models, we simply consider a configuration 
containing exactly one excitation,
which has escape rate $2dc$ in the
FA case and $dc$ in the East model, where $d$ is the spatial
dimension (in the bosonic variants, these rates are $2de^{-\beta}$ and
$d e^{-\beta}$).  In the (2)-TLG, all of the particles in the model
can be arranged in a single compact cluster, in which all but six
of the particles are unable to move: this configuration has
$r(\C)=6$.   For the KA model, a similar construction leads to
configurations with $r(\C)=4$.  
Thus, combining~(\ref{eq:bound_zero}-\ref{eq:kcm_r}), we have
\begin{equation}
\lim_{\vol \to\infty} \frac 1\vol  \psi_K(s)=0, \qquad s\geq0
\end{equation}

\begin{figure}[t]
  \centering
  \includegraphics[width=0.28\columnwidth]{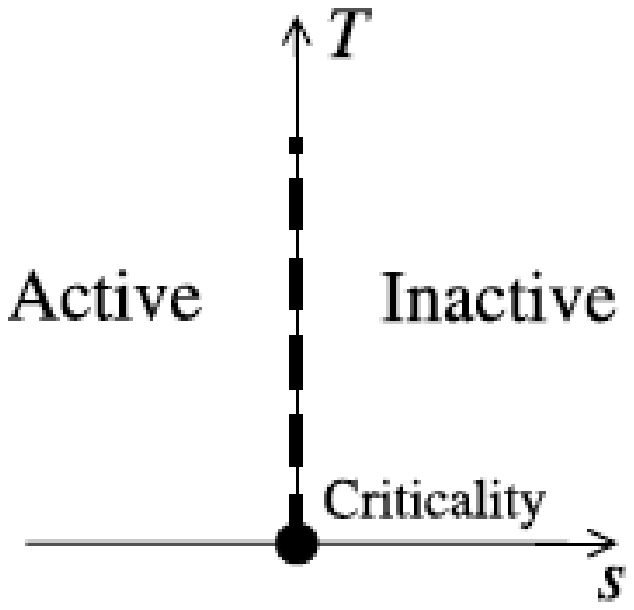} 
  \includegraphics[width=0.34\columnwidth]{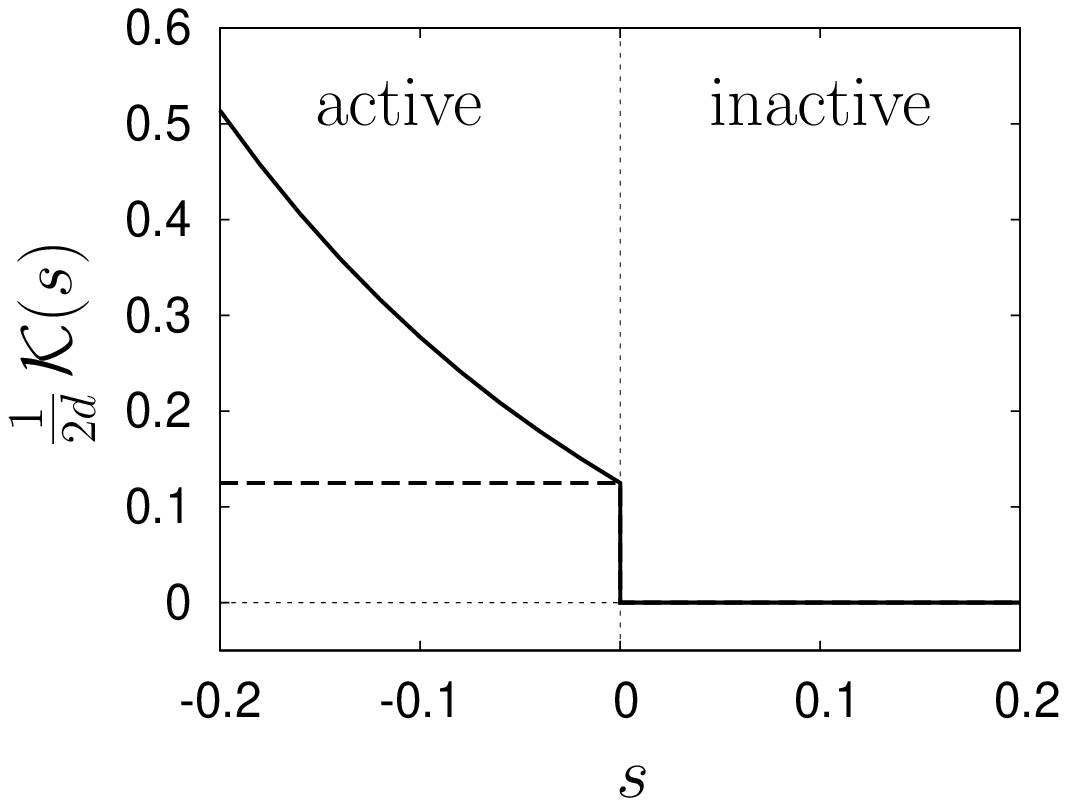} \quad
  \includegraphics[width=0.34\columnwidth]{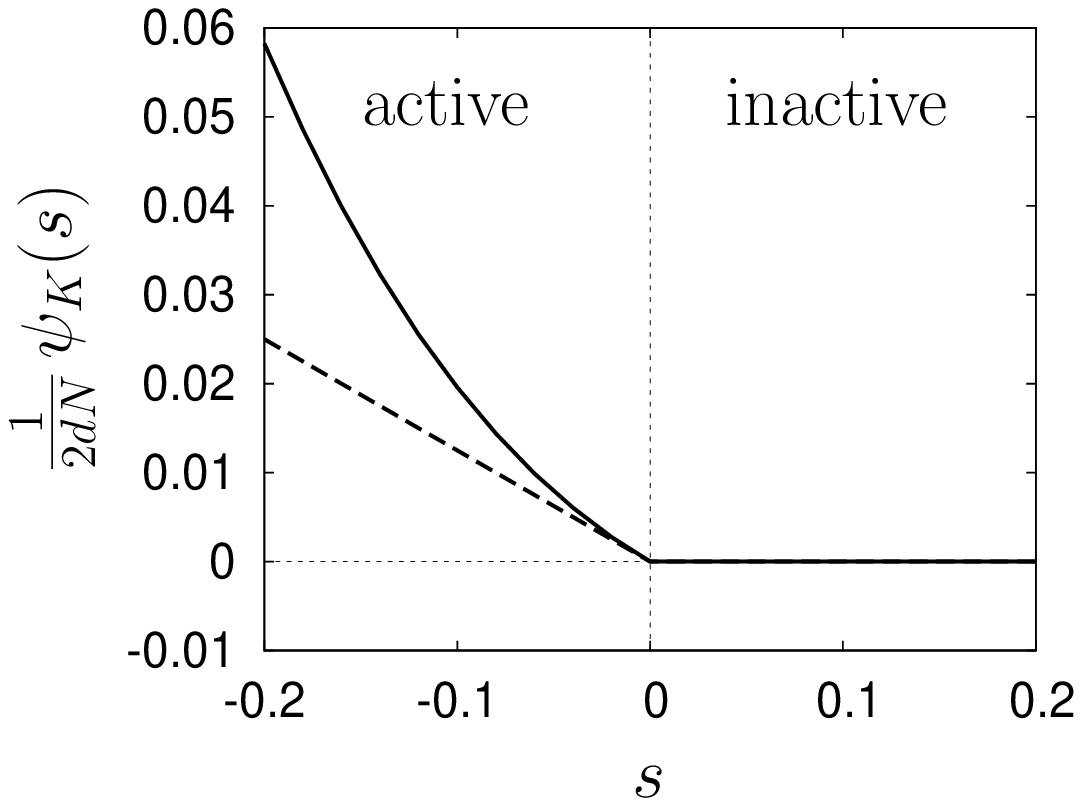} \quad
  \caption{ 
(Left): Generic `dynamic phase diagram' for 
spin-facilitated KCMs such as the FA and East models.  There is a dynamical
phase coexistence boundary at $s=0$,  for all finite temperatures.  The boundary
ends in a dynamical critical point at $s=T=0$.  For the (2)-TLG
and KA models, the picture is identical if the temperature $T$
is replaced by the fraction of vacant sites, $1-\rho$.
(Center): Variational estimates for 
the activity per site $\mathcal{K}(s)$,
in the bosonic FA model, at $c=0.25$.  For $s>0$, the result $\mathcal{K}(s)=0$
is exact in all dimensions, from Eq.~\eqref{eq:Ks_zero}.  For $s<0$,
the dashed line shows the lower bound obtained from~\eqref{eq:Ks_finite},
while the solid line is the variational estimate $
\mathcal{K}_\mathrm{var}(s)=-\frac 1\vol (\mathrm{d}/\mathrm{d}s)\psi_\mathrm{var}(s)$,
obtained from~\eqref{eq:psiK_FA_MF}.  As discussed in the text, the solid
line gives the exact result for the mean-field variant of the FA model.
(Right) Again, we show the exact result $\psi_K(s)=0$ for $s>0$, together
with the variational lower bounds~\eqref{eq:psiK_FA_MF} (solid line, exact
for the mean-field variant)
and~\eqref{eq:bound_r} (dashed line).
}
  \label{fig:FA_transition_bounds}
\end{figure}

Recalling that $\langle K\rangle=t(\mathrm{d}/\mathrm{d}s)\psi_K(s)$, We define the mean
activity per site per unit time as 
\begin{equation}
\mathcal{K}(s)= \lim_{\vol \to\infty}\frac 1\vol \frac{\mathrm d \psi_K(s)}{
\mathrm d s}
\end{equation}
and we can see that
\begin{equation}
\label{eq:Ks_zero}
\mathcal{K}(s)=0, \qquad s>0
\end{equation}
Further, from Eq.~\eqref{eq:K-R-mean}, we have
$K(0)=t\langle r\rangle=
  t\sum_\C P_\mathrm{eq}(\C) r(\C)$.
Since the distributions
$P_\mathrm{eq}(\C)$ have simple forms in kinetically
constrained models, this quantity can be calculated explicitly:
the limit 
$\mathcal{K}_\mathrm{eq}=
\lim_{\vol \to\infty} \vol ^{-1} \sum_\C P_\mathrm{eq}(\C) r(\C)$ is
finite and positive for all the models that we consider.  
Finally, it follows
from~\eqref{eq:def_Z_A_Laplace} 
that $\mathcal{K}(s)$ is non-increasing, so 
that
\begin{equation}
\label{eq:Ks_finite}
\mathcal{K}(s) \geq \mathcal{K}_\mathrm{eq}, \qquad s\leq0
\end{equation}
with $\mathcal{K}_\mathrm{eq}$ finite.  
Eqs.~\eqref{eq:Ks_zero} and~\eqref{eq:Ks_finite} 
establish the discontinuity of
$\mathcal{K}(s)$ at $s=0$: in the 
limit of large system size, the dynamical
free energy has a discontinuous first derivative which
we refer to as a first-order dynamical phase transition.
We have established the existence of such a transition
in the FA, East and (2)-TLG models, in all dimensions
and for all finite temperatures [and for
all finite densities $\rho$ in the (2)-TLG].  That is,
the simple phase diagram shown in fig~\ref{fig:FA_transition_bounds}
is generic to all of these models.  

\subsection{Variational free energy for the excitation density $\rho_K(s)$}
\label{sec:bosonic_FA_var}

The analysis given above establishes some minimal
conditions that are sufficient for the existence of a 
first-order transition.  For a more quantitative analysis,
it is useful to use a specific variational distribution
in~\eqref{eq:max_P}.  We consider a general bosonic KCM with single
spin-flip dynamics, and we define
a distribution of the excitation numbers $n_i$ that
is independent of the site $i$, and parameterized by
a mean density $\rho$:
\begin{equation} \label{eq:V_rho}
V_\rho(\{n_i\}) = \prod_i \sqrt{\frac{\rho^{n_i} e^{-\rho}}{n_i!}}
\end{equation}
From~\eqref{eq:max_P}, we therefore have
$
\psi(s) \geq -\vol  \min_\rho \mathcal{F}_K(\rho,s)
$
with 
\begin{equation} \label{eq:def_F_rho}
 \mathcal{F}_K(\rho,s) \equiv \vol ^{-1} 
    \frac{ \langle V_\rho| \tilde{\mathbb W_K}|V_\rho\rangle}
                            {\langle V_\rho|V_\rho\rangle}
\end{equation}
The value of $\rho$ which minimises $\mathcal{F}_K(\rho,s)$ is denoted
by $\rho_\mathrm{var}(s)$.  It represents a variational estimate
for the order parameter $\rho_K(s)$: if the variational 
bound~\eqref{eq:max_P} is saturated then $|V\rangle$
is an eigenvector of the symmetrised operator $\tilde{\mathbb W}_K$,
and it follows from~\eqref{eq:ave_B_V} that 
$
\rho_K(s)=\rho_\mathrm{var}(s)/(1-\ee^{-\rho\vol}).
$

\begin{figure}[t]
  \centering
  \includegraphics[height=4cm]{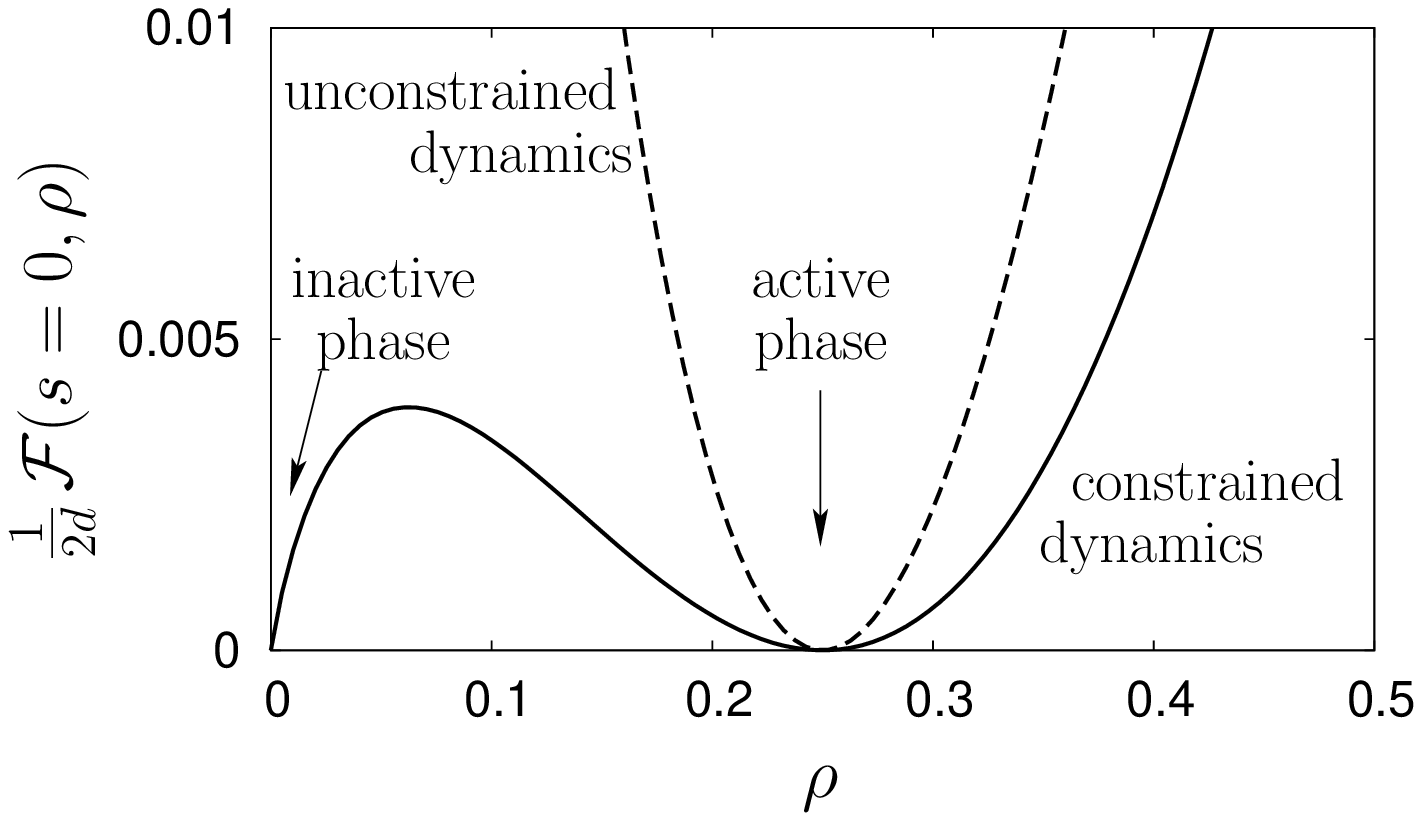}  
  \includegraphics[height=4cm]{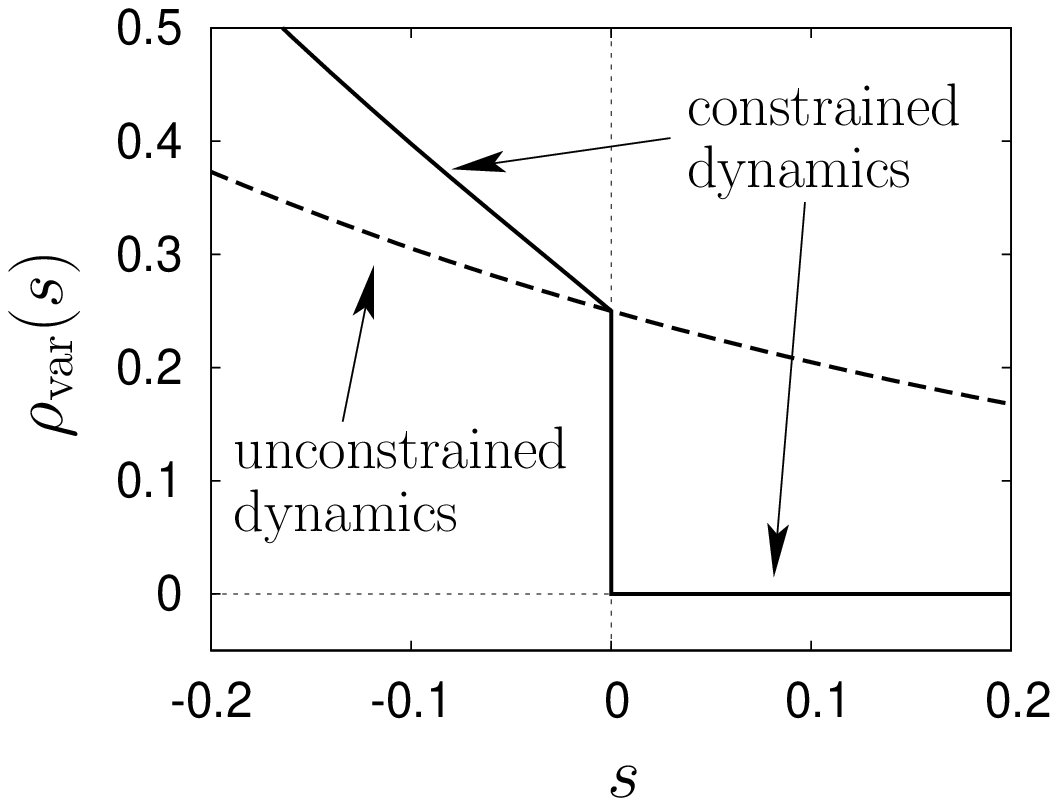}  
  \includegraphics[width=0.4\columnwidth]{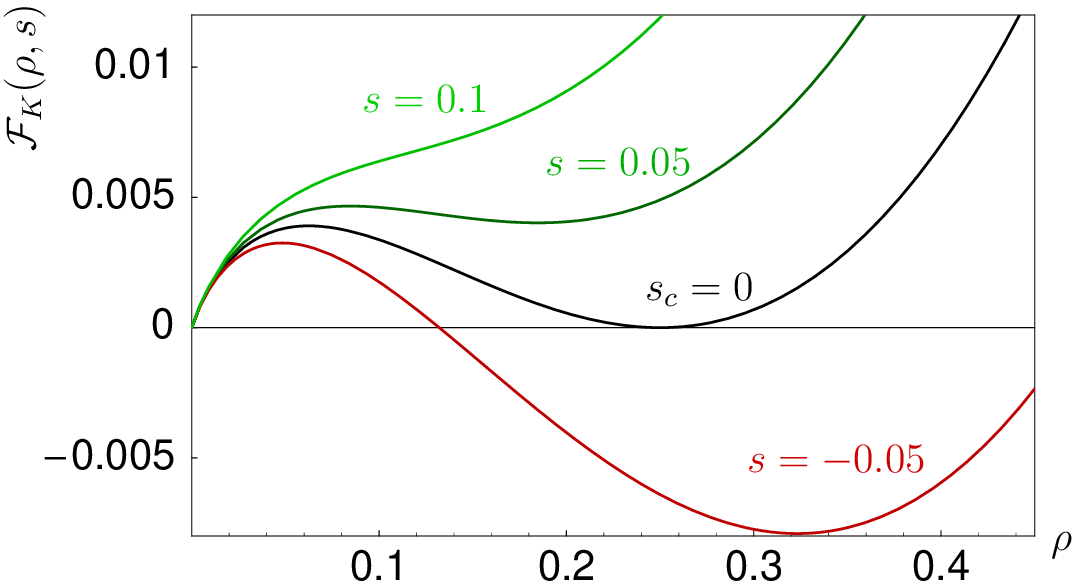}  
\caption{
(Top~left)~Variational
free-energies for the FA model (constrained dynamics, solid line),
and the A-model (unconstrained dynamics, dashed line) at $s=0$.  
Both models have the same thermodynamic free energy.  However, the 
dynamical function $\mathcal{F}(\rho,s)$ 
reveals that the FA model has two dynamical phases
while the A-model has only an active phase.  
(Top~right)~The variational
estimate $\rho_\mathrm{var}(s)$, for the FA model (solid line) and the A-model (dashed).  
For the mean-field FA model, $\rho_\mathrm{var}(s)$
coincides with $\rho_K(s)$ in the limit of large system size $N$; 
for the A-model, 
$\rho_K(s)=\rho_\mathrm{var}(s)$ always.
(Bottom)~Dependence
of the variational free energy on the field $s$, in the FA model.  At
the phase coexistence condition, $s=0$, the free energy has degenerate
minima.  For finite $s$, either the inactive or active phase is preferred.
}
\label{fig:FreeEn_FA_uncons}
\end{figure}

For the bosonic FA model, it is straightforward to calculate
$\mathcal{F}_K(\rho,s)$.  
The only subtlety is that we must
explicitly exclude the state with no excitations from the inner 
products, as discussed in Section~\ref{sec:reducibility}. 
In the Doi-Peliti formalism, our choice for $V(\{n_i\})$
renders this calculation very simple: in terms of the
symmetrised operator $\tilde{\mathbb W}_K^\mathrm{(FA)}$, we have
\begin{equation}
  \mathcal F_{K}(\rho,s) = \vol ^{-1} \frac{ 
\langle 0 | \ee^{-\sqrt{\rho}\sum_i a_i}
  \tilde{\mathbb W}_K^\mathrm{(FA)} \ee^{-\sqrt{\rho}\sum_i a_i^\dag}|0\rangle 
\ee^{-\rho N}
}
 { 1 - e^{-\rho N}}.
\end{equation}
Hence,
\begin{equation} \label{eq:F_rho_s_FA}
  \mathcal F_{K}(\rho,s) = 2d\frac{
    c+\rho - 2 \ee^{-s}\sqrt{c\rho}}
 {1-\ee^{-\rho N}}\rho.
\end{equation}
Minimising over $\rho$, we find $\lim_{\vol \to\infty}\vol ^{-1}\psi(s)
  \geq\psi_\mathrm{var}(s)$ with
\begin{equation}
 \label{eq:psiK_FA_MF}
  \psi_\mathrm{var}(s)= 
          \frac{2d}{3} \rho_\mathrm{var}(s)[\rho_\mathrm{var}(s)-c]
\end{equation}
and
\begin{align}
 \label{eq:rhoK_FA_MF}
  \rho_\mathrm{var}(s)&=\left\{
    \begin{array}{ll}
      0, & \quad  s>0 \\
       (c/8) \Big(9\ee^{-2s}-4+3\ee^{-s}\sqrt{9\ee^{-2s}-8}\Big),
       & \quad s\leq0
    \end{array}
    \right.
\end{align}
(Within this approach, we obtain $\rho_\mathrm{var}(0)=c$
by minimising $\mathcal{F}(\rho,s)$ at fixed system size
$N$, and then taking $N\to\infty$.)
The bound on $\psi_K(s)$ and the corresponding estimate
of $\mathcal{K}(s)$ are shown fig~\ref{fig:FA_transition_bounds}.  
The variational estimate for $\rho_K(s)$ and the variational
free energy $\mathcal{F}_K(\rho,s)$ are shown in 
Fig~\ref{fig:FreeEn_FA_uncons}.

So far, we have used Eq.~\eqref{eq:max_P} to obtain variational estimates for
$\psi_K(s)$ and $\rho_K(s)$ for the FA model
in finite dimension.  For the mean-field variant
of the FA model, it can be shown that these variational
estimates are exact, in the limit of large system size $N$.  (The
factor $2d$ that appears in $\mathcal{F}(\rho,s)$ is simply an
arbitrary rescaling of time in the mean-field model.  Our definition
of the mean-field model  requires that we set $2d=1$.)
That is, the difference between the variational ansatz of~\eqref{eq:V_rho}
and the dominant eigenvector of $\tilde{\mathbb W}_K$ vanishes
at large $N$.  Mean-field models are discussed in more
detail in Section~\ref{sec:traj} below.

It is useful to compare these results for the FA model with
the bosonic variant of the A-model,
for which it can be easily verified that
the large deviation function $\psi_K(s)$
coincides with the variational bound $\psi_\mathrm{var}(s)$,
even for finite system size $\vol $.  In that case, we have 
\begin{equation}
\mathcal{F}(\rho,s)=c+\rho-2\ee^{-s}\sqrt{\rho c}\qquad  
  \psi_K(s)=c(e^{-2s}-1),\qquad \rho_K(s)=c\ee^{-2s}
\end{equation}
Thus, while
constrained FA model and the unconstrained A-model possess 
the same equilibrium distribution $P_\mathrm{eq}(\C)$, and hence the
same \emph{static} free energies, 
their {dynamical} free energies show dramatic differences. For
large systems, the FA model exhibits a dynamical phase transition,
while the A-model does not.  See fig.~\ref{fig:FreeEn_FA_uncons}.

The presence of the dynamical first-order transition in the FA
model is intimately connected to the two minima in 
$\mathcal{F}_K(\rho,s)$.  As shown in Fig.~\ref{fig:FreeEn_FA_uncons},
these two minima represent an active phase,
with $\rho\simeq c$ and an inactive one with $\rho\simeq0$.  
The global minimum of $\mathcal{F}(\rho,s)$ lies in the active
phase for $s<0$, while it lies in the inactive phase for $s>0$.  
For $s=0$, one must consider carefully the limit of large
system size $\vol $: we have $\mathcal{F}_K(c,0)=0$ while the inactive
minimum occurs at $\rho_\mathrm{var}=\mathcal{O}(\vol ^{-2})$, 
where the value of the variational bound $\psi_\mathrm{var}$ is 
positive.  Thus, the global minimum of $\mathcal{F}_K(\rho,s=0)$
occurs at 
the active state density.  However, any $s>0$ is sufficient to drive
the system into the inactive phase.
The effect arises because of two non-commuting
limits: when minimising $\mathcal{F}(\rho,s)$,
taking the limit $s\to 0$ before the limit of large $\vol $ results
in active behaviour; on the other hand, taking the limit of large $\vol$ 
followed by a limit $s\to0^+$ leads to the inactive phase.  

We note that
the dynamical phase transition in the FA model requires a limit
of large system size ($\vol \to\infty$)
as well as a limit of long trajectories 
($t\to\infty$).  To keep our methods well-defined, 
we excluded the configuration with no excitations from
the initial conditions (recall Section~\ref{sec:reducibility}).  
We emphasise that we have proved the existence
of a dynamical phase transition in an irreducible model, with
no absorbing states (this can be compared, for example, with
phase transitions in the directed percolation universality
class~\cite{HinrichsenDP}).

\subsection{Numerical results}
\label{sec:numericalmethod}

\subsubsection{Cloning method}
\label{sec:clones}

We now present some numerical computations of the dynamical free
energy $\psi_K(s)$ in KCMs.  From~\eqref{eq:def_Z_A_Laplace}, this can
be obtained from the large $t$ limit of the equilibrium average
$\langle\ee^{-sK}\rangle$.  However, direct calculation of this
average requires a computational effort that scales exponentially with
$t$: the average is dominated by rare histories lying in the tails of
the distribution of $K$. 
In dynamical systems~\cite{tailleurNPhys} 
and in discrete time Markov processes~\cite{Giardina},
this problem can be avoided by using a cloning method similar to that used 
in quantum mechanical Diffusion Monte Carlo algorithms~\cite{DiffusionMC}.  
This method was generalised to continuous time Markov processes 
by Tailleur and Lecomte~\cite{clones}.
Here, we briefly summarize the
algorithm for obtaining dynamical free energies.

The function $\psi_A(s)$ is obtained as the largest eigenvalue of 
the operator $\mathbb W_A$.  However,
this operator does not conserve probability 
[that is, $\mathbb W_A$ sets the time
dependence of $\hat P_A(\C,s,t)$, but the
`total probability' $\sum_\C \hat P_A(\C,s,t)$ is not
a constant of the motion, except at $s=0$].
To interpret this non-conservation, we define a new
stochastic process (a `modified dynamics') with rates
$W_s(\C'\to\C)$, chosen so that we can decompose
%
\eqref{eqn:evol_hatP} as
\begin{equation}\label{eq:decomp_dPdt_clones}
     \partial_t  \hat P_A(\C,s,t) = 
         \sum_{\C'} \left[ W_s(\C\to\C') - r_s(\C)\delta_{\C,\C'} \right] 
  \hat P_A(\C',s,t) 
        +  {\delta r_{s}}(\C)\hat P_A(\C,s,t)
\end{equation}
with $r_{s}(\C)=\sum_{\C'}{ W_{s}}(\C\to\C')$
and $\delta r_s(\C)=r_s(\C)-r(\C)$.
This decomposition is discussed in appendix~\ref{app:AB},
and the rates $W_s(\C\to\C')$ are given in~\eqref{eq:rates_Ws}.

For the purposes of the cloning algorithm, we note that the
first term in~\eqref{eq:decomp_dPdt_clones} conserves
probability (in the sense given above),
while the second term
represents the creation or destruction of copies (clones) of the system.
That is, starting from a
large number of copies of the system, we let each copy of the system
evolve with the modified dynamics (rates $W_s$). In addition, the
copies are subject to a creation/destruction process with a
configuration-dependent rate $\delta
r_{s}(\C)$.  In this way, the number $n_\mathrm{cl}(\C,t)$ of copies of
the system in configuration $\C$ at time $t$ has the same time
evolution as $\hat P_A(\C,s,t)$ in~\eqref{eq:decomp_dPdt_clones}.  To
avoid the ensuing exponential increase or decrease of the total
number of copies [which behaves as $\ee^{t\psi_A(s)}$], one
compensates the clone creation/destruction rates 
of~\eqref{eq:decomp_dPdt_clones}  with
configuration-independent creation/destruction rates.  The rates
are adapted as the simulation proceeds, in order to keep a constant clone 
population~\cite{Giardina,clones}.  These adaptively determined
rates can then be used to 
obtain the dynamical free energy $\psi_A(s)$.

\subsubsection{Results}

\begin{figure}[t]
  \centering
  \includegraphics[width=0.39\columnwidth]{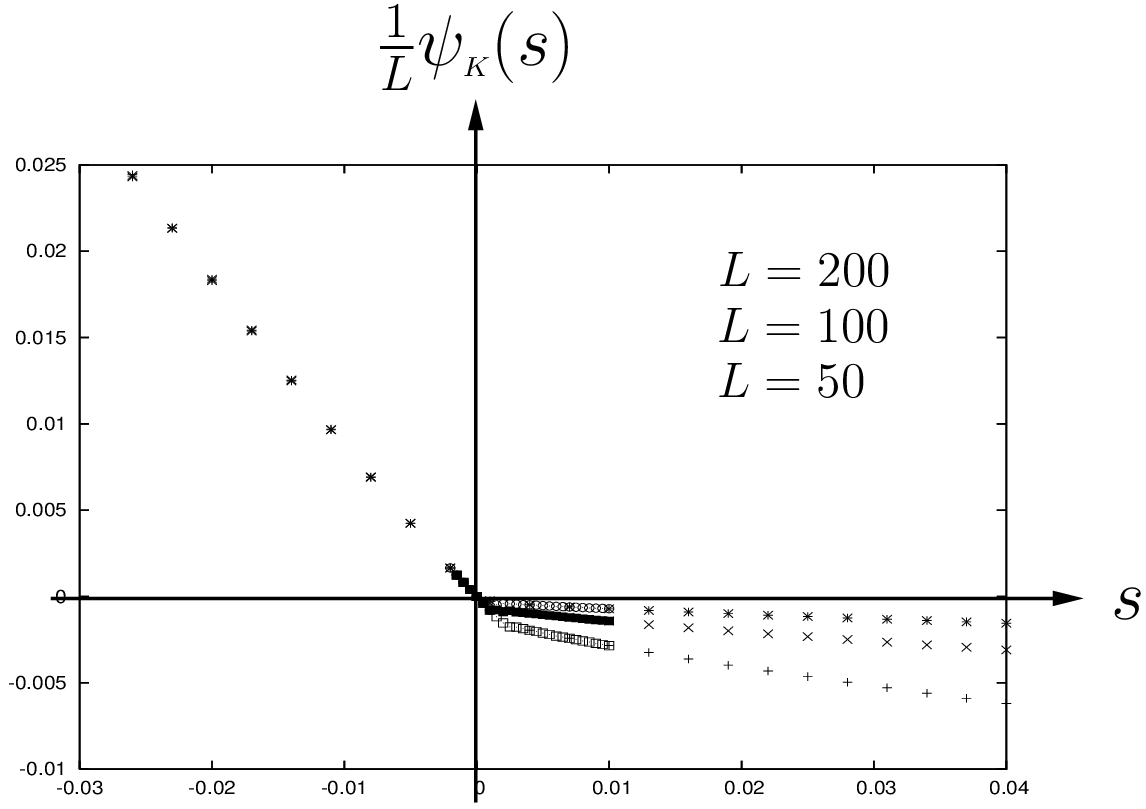} \qquad
  \includegraphics[width=0.39\columnwidth]{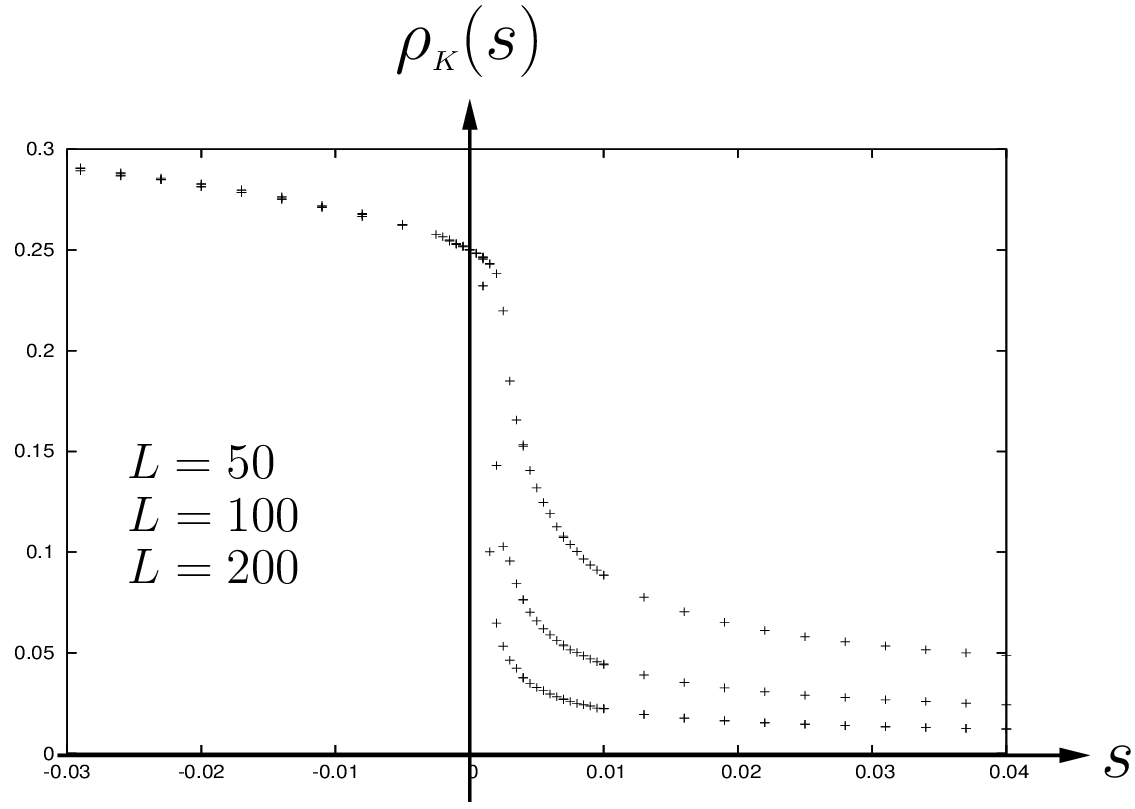}
  \includegraphics[width=0.39\columnwidth]{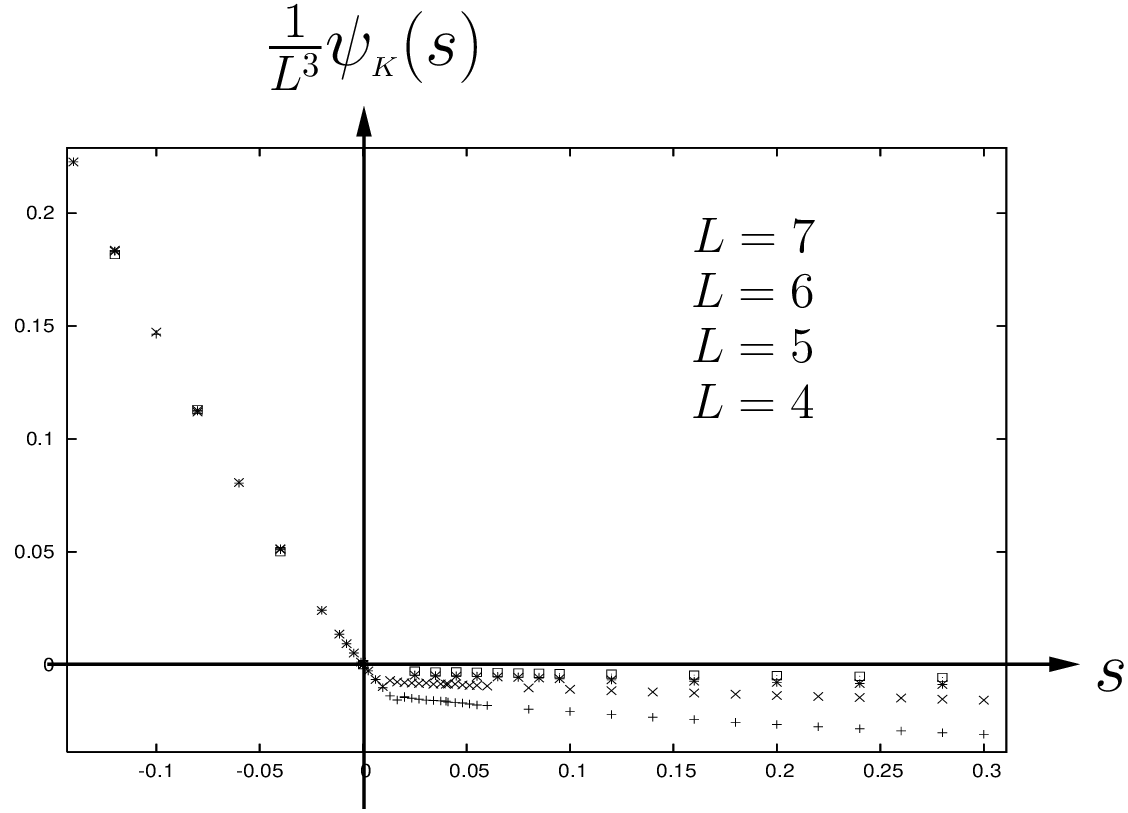} \qquad
  \includegraphics[width=0.39\columnwidth]{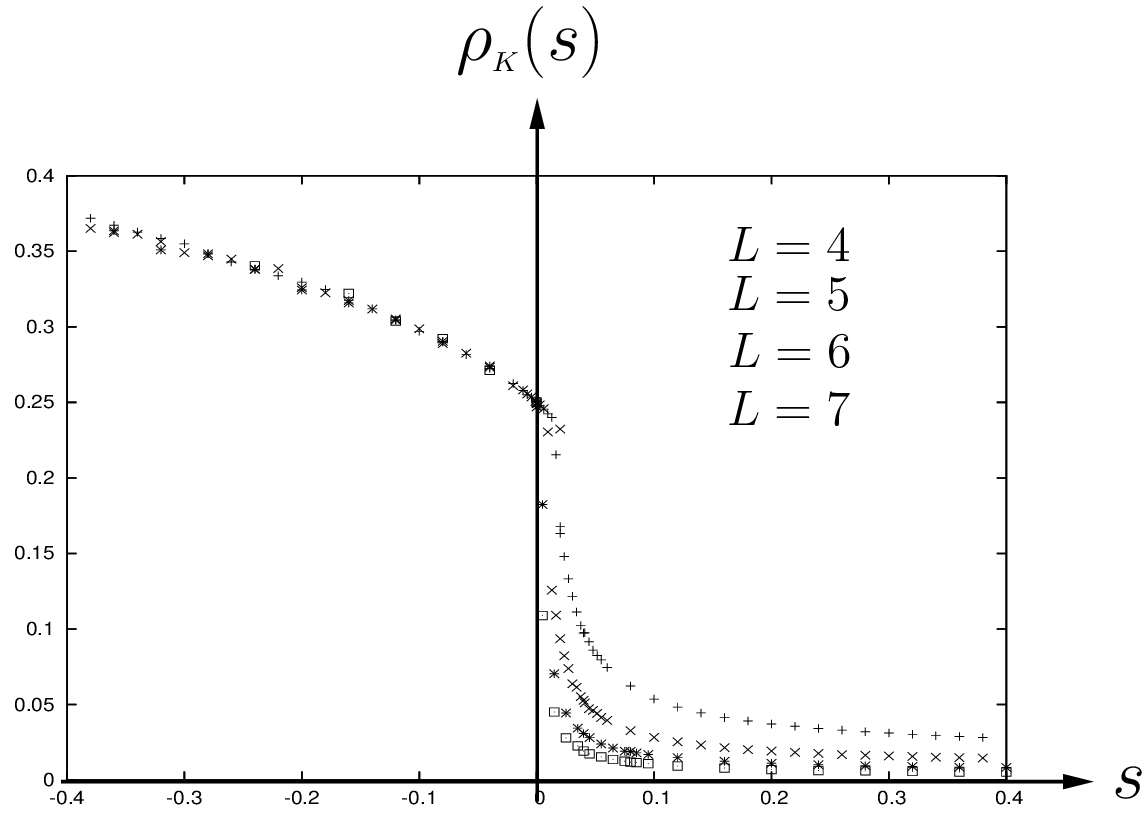}
  \caption{Finite-size scaling in the $s$-ensemble  associated to $K$,
    in the vicinity of the singular point $s_{c}=0$ for KCMs with non-conserved number of particles:
    Fredrickson-Andersen model in dimension $d=1$ 
    \textbf{(top)} and East model in dimension $d=3$ \textbf{(bottom)}. The temperature is $T=1/\beta=0.91$, and linear system sizes $L$ are
    given on the graphs.    The finite-size scaling    illustrates the first order dynamical phase transition in  $s_{c}=0$.
    \textbf{(Left)} large deviation function $\frac 1 {L^d}
    \psi_{K}(s)$. \textbf{(Right)} density of excited sites
    $\rho_{K}(s)$.}
  \label{fig:FA_EAST}
\end{figure}

\begin{figure}[t]
  \centering
  \includegraphics[width=0.39\columnwidth]{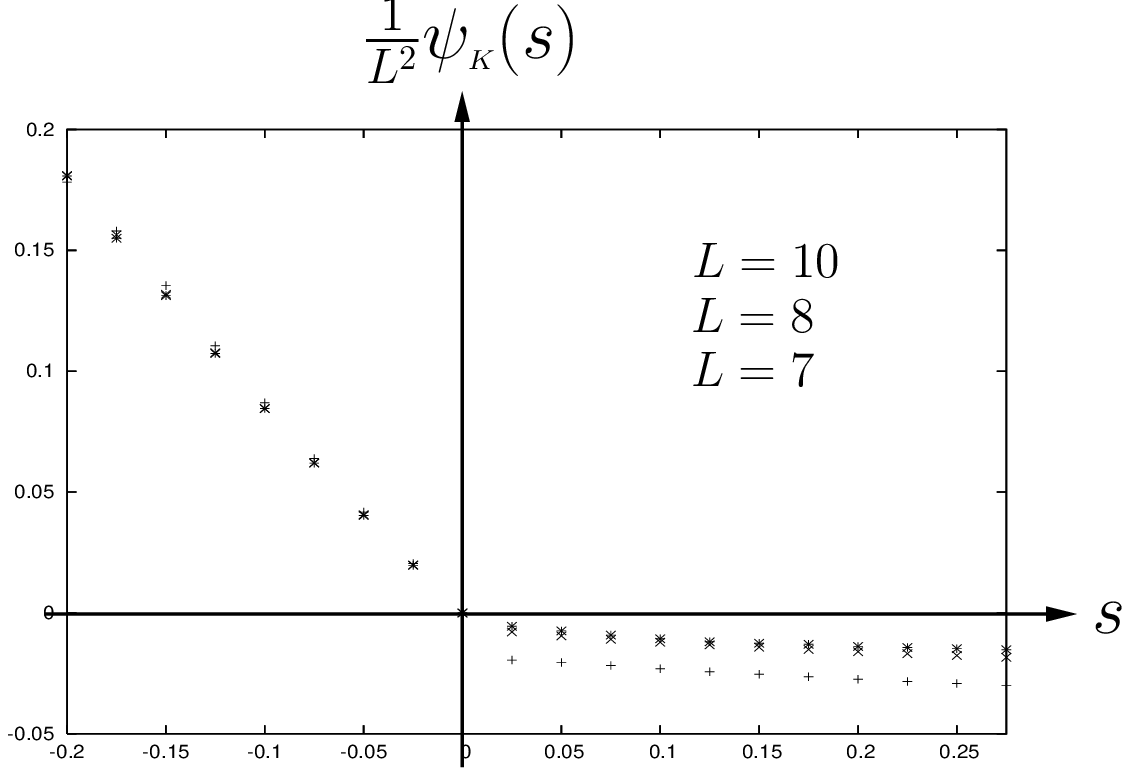} \qquad
  \includegraphics[width=0.39\columnwidth]{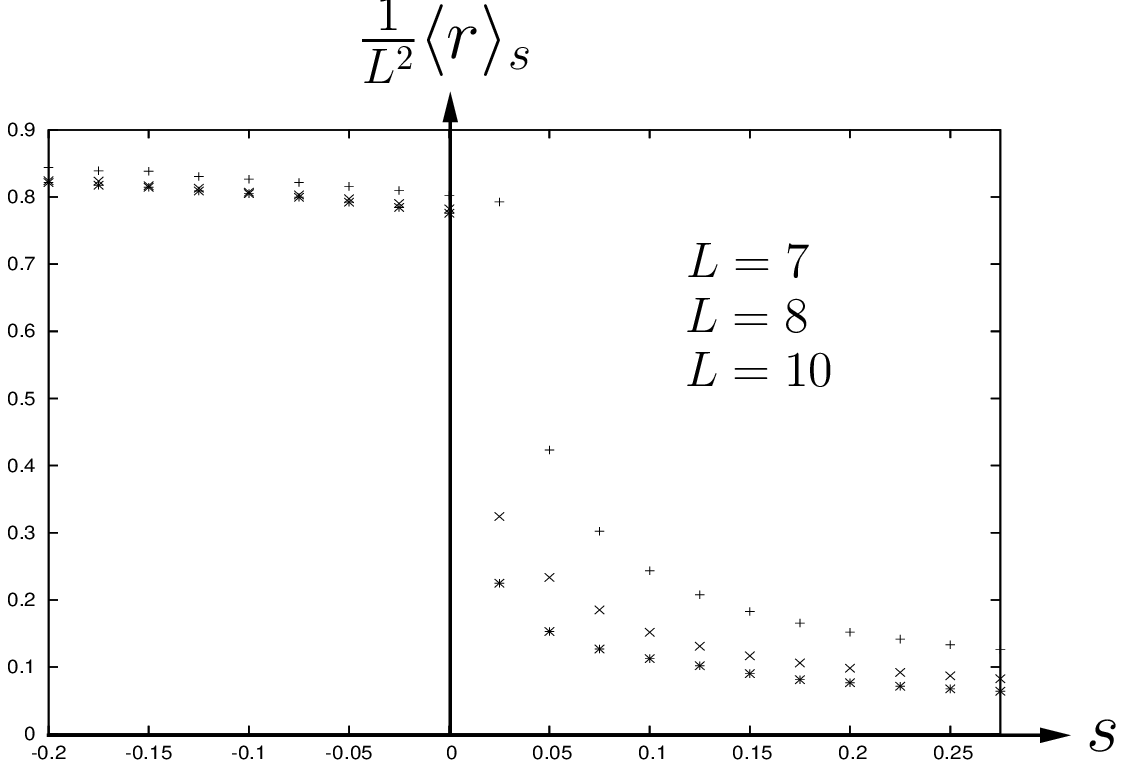}
  \includegraphics[width=0.39\columnwidth]{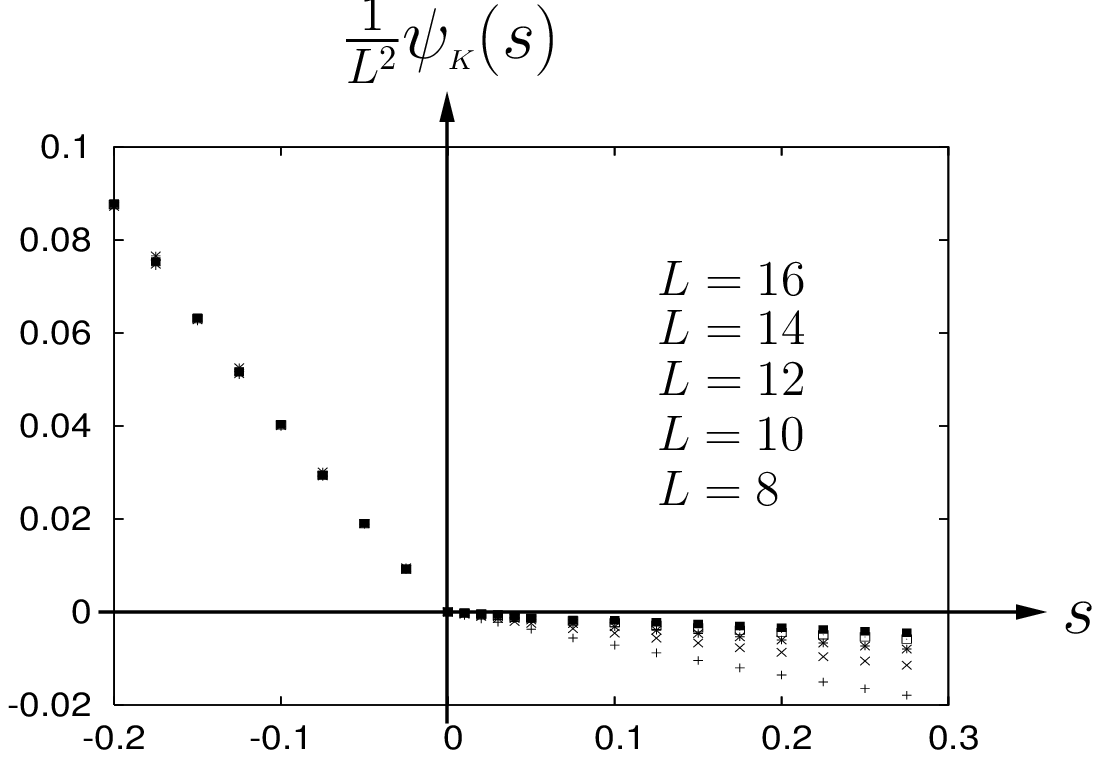} \qquad
  \includegraphics[width=0.39\columnwidth]{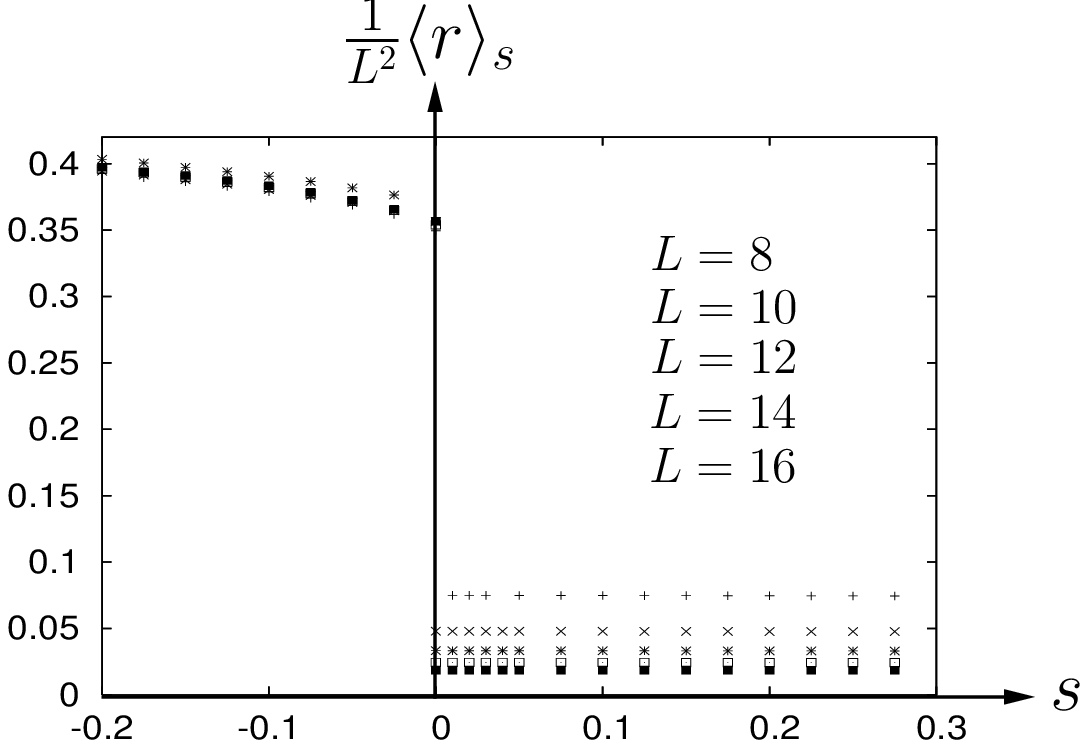}
  \caption{Finite-size scaling in the $s$-ensemble associated to $K$,
    in the vicinity of the singular point $s_{c}=0$ for KCMs with
    conservation of particle number: Kob-Andersen model
    \textbf{(top)} and Triangular Lattice Gas model \textbf{(bottom)}
    Particle density is $\rho=0.5$, and linear system sizes $L$ are given on the graphs.  The
    finite-size scaling illustrates the first order dynamical phase
    transition in $s_{c}=0$.
    \textbf{(Left)} large deviation function $\frac 1 {L^d}
    \psi_{K}(s)$. \textbf{(Right)} order parameter $\frac 1{L^d}
    \langle r\rangle_s$.}
  \label{fig:KA_TLG}
\end{figure}

Using the cloning method, we investigated two classes of KCMs.
  In Fig.~\ref{fig:FA_EAST} we consider spin-facilitated models:
  the FA model in one dimension and the East model in three
  dimensions.  We evaluated the free energy
  density $\frac 1 \vol  \psi_K(s)$ for various system sizes. Its
  behavior as a function of $\vol $ drastically depends on the sign of
  $s$, as is also the case for the order parameter $\rho_K(s)$.
  Negative values of $s$ correspond to active histories, with non-zero
  mean density of particles, while for positive values of $s$, the
  mean number of particle in the system remains finite, leading to a
  zero density and activity in the infinite system size limit.

In Fig.~\ref{fig:KA_TLG}, we consider two models with
particle conservation: the
   KA and (2)-TLG models, both in two
   dimensions. Remarkably, the picture is very similar to the previous one,
   the (conserved) density being replaced with the order
   parameter $r_K(s)$.

In the active phase ($s<0$), the order parameters $\rho_K(s)$  and $r_K(s)$
converge rapidly as the system size $N$ is increased, for all the KCMs
that we considered.
On the other hand, in the inactive phase ($s>0$)
the order parameters decrease with system size as $N^{-1}$.
Comparing Fig.~\ref{fig:FA_EAST} 
and Fig.~\ref{fig:FA_transition_bounds} confirms the analysis
based on variational bounds on $\psi_K(s)$: in the limit
of large system size, KCMs exhibit
dynamical first-order transitions at $s=0$.  For models on finite
lattices, the equilibrium ($s=0$) dynamics are representative
of the active phase, and the system crosses
over to the inactive phase at a value of $s$ that scales
as $N^{-1}$ for large $N$.

\subsection{Criticality at zero temperature and dynamical phase transition}
\label{sec:AAmodel}

We emphasize that while a zero temperature dynamical 
critical point is
common to many KCMs~\cite{Whitelam}, this is not a sufficient 
condition for dynamical
phase coexistence.  Rather, the relevant feature is
the presence of states with subextensive escape rates,
as discussed in Section~\ref{sec:phasecoexist}.
In this section, we consider the
AA model.  The FA and AA models both have zero-temperature 
dynamical critical points, with the same scaling
exponents and closely related correlation functions~\cite{Jack-Mayer}. 
However, all states in the AA model have extensive escape rates, 
so we do not expect any transition at $s=0$. In the following, 
we show that this is indeed the case, by discussing the
AA model both within a mean field approximation and 
using exact results in one dimension.

\subsubsection{`Mean-field' variational bound}

We consider the bosonic AA model in dimension $d$.
Following Section~\ref{sec:bosonic_FA_var}, 
we calculate the variational Landau free energy using the
Doi-Peliti representation, obtaining
\begin{align}
  \mathcal F_K(\rho,s)&= 2d
  \big[  2D \rho(1-\ee^{-s}) +k'+k \rho^2 - 2\ee^{-s}\rho\sqrt{k k'}\big]
\end{align}
and we identify $\psi_\mathrm{var}(s)= -\min_\rho F_K[\rho,s]$
as a lower bound on $\vol \psi_K(s)$.

The variational estimate for the steady-state density and
the variational bound are
\begin{align}
  \rho_\mathrm{var}(s) &= \ee^{-s}\sqrt{\frac{k'}{k}} + 
                             (\ee^{-s}-1)\frac{D}{k} 
\\
  \psi_\mathrm{var}(s) &= 2d \left(k\rho^2_\mathrm{var}(s)-k'\right)
\end{align}
The variational bound is is indeed analytic for all $s$, consistent
with our intuition that the AA model has no dynamical
phase transition.  Again, 
these variational estimates are exact for the mean-field
AA model in the limit of large system size, if we set $2d=1$.

\subsubsection{AA model in one dimension}

\begin{figure}[t]
  \centering
  \includegraphics[width=0.49\columnwidth]{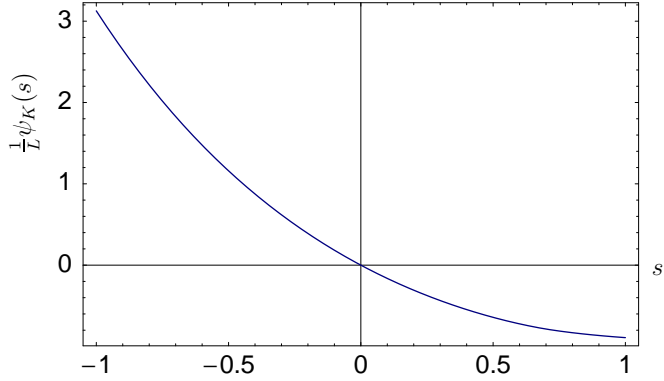}
  \caption{ Dynamical free energy density $\frac 1 {\vol }\psi_{K}(s)$ for
    the AA model in dimension one. It presents no singularity at
    $s=0$.  Although the AA model can be mapped to the FA model and
    displays the same critical properties at zero temperature, it
    is not subject to dynamical phase coexistence.  }
  \label{fig:AA_psiK}
\end{figure}

In addition to the mean-field case, we can also obtain
the large deviations of the AA model in one dimension,
through a mapping to a free fermion system.
The evolution operator associated to $K$
for AA model can be written in a spin-$\frac{1}{2}$ representation
(recall Eq.~\eqref{eq:dynAA} and see, for example, Ref.~\cite{StinchReview}):
\begin{align}
   \mathbb W_K = 
   \sum_i &  \Big\{z\big[ k \sigma_i^-\sigma_{i+1}^-+k' \sigma_i^+\sigma_{i+1}^+ + D( \sigma_i^- \sigma_{i+1}^++ \sigma_i^+\sigma_{i+1}^-)\big]
\nonumber \\ &
   -k\hat n_i \hat n_{i+1} -k'(1-\hat n_i)(1-\hat n_{i+1})- D \hat n_i (1-\hat n_{i+1})- D \hat n_{i+1} (1-\hat n_{i})\Big\}
\end{align}
where $z=\ee^{-s}$, $\sigma_i^\pm = 
 \frac{1}{2}(\sigma_i^x\pm{\rm i}\sigma_i^y)$, 
$\hat n_i = \frac{1+\sigma_i^z}{2}$ and $  \sigma_i^x, \sigma_i^y,  \sigma_i^z $
are the usual Pauli matrices. In the spin language, the presence (or absence) of a particle at site $i$ is coded by an up (or down) spin. We 
use the detailed balance property to symmetrise this operator 
followed by a
Jordan-Wigner transformation~\cite{Jordan-Wigner,StinchReview}
\begin{equation}
  \sigma_i^+ = c_i \exp\Big(i\pi \sum_{j=1}^{i-1} c_j^\dag c_j\Big)
\qquad
  \sigma_i^- =  \exp\Big(i\pi \sum_{j=1}^{i-1} c_j^\dag c_j\Big) c_i^\dag
\end{equation}
which allows us to represent the spin operators in terms of fermionic
creation/annihilation operators $c_j^\dag$ and $c_j$.  For values of
the parameters verifying $k+k'=2D$~\cite{freeF}, this puts $\tilde{\mathbb
  W}_K$ into a quadratic form:
\begin{equation}
\tilde{\mathbb W}_K=-\sum_q\left[(k-k'-z(k+k')\cos q)c_q^\dagger c_q
-iz\sqrt{kk'}c^\dagger_q c^\dagger_{-q} \sin q
+iz\sqrt{kk'}c_{-q} c_q\sin q\right]-k'\vol 
\end{equation}
where we introduced Fourier-transformed
operators 
$c_q=\sum_j c_j e^{iqj}$ and $c_q^\dagger=\sum_j c_j^\dagger e^{-iqj}$.

We now introduce new fermionic operators
$\beta_q = \cos\theta_q c_q - i \sin\theta_q c^\dagger_{-q}$,
$\beta_q^\dagger = \cos\theta_q c^\dagger_q + i \sin\theta_q c_{-q}$.
Taking $\pi/4<\theta_q<\pi/4$, we write
$\sin2\theta_q=(2z\sqrt{k k'}\sin q/\Omega_q)$, with
\begin{equation}
\Omega_q=\sqrt{4kk'(z^2-1)+[k+k'-z(k-k')\cos q]^2}
\end{equation}
Then, the dynamical free energy is the largest eigenvalue of the operator
\begin{equation}
\tilde{\mathbb W}_K = \frac 1 2
\sum_q \left[ \Omega_q (1 - 2 \beta_q^\dagger \beta_q) - (k+k') \right]
\end{equation}
Finally, for large $N$, we convert 
the sum over $q$ to an integral, arriving at
\begin{equation}
\psi_K(s)=
\frac{\vol}{2}\left[ -(k+k') +
\int\frac{\dd q}{2\pi}\, \Omega_q  \right]
\end{equation}
which depends on $s$ through the dependence of $\Omega_q$ 
on $z=\ee^{-s}$.

This exact result for the AA model in $d=1$ shows that
the large deviation function $\psi_K(s)$ is analytic at $s=0$ (see
also Fig.~\ref{fig:AA_psiK}), as opposed to the one of the FA model.
Despite the presence of a dynamical
critical point, the AA model has no configurations with subextensive
escape rate $r(\C)$, and does not exhibit dynamical phase coexistence
(in the vicinity of $s=0$).

\section{Properties of trajectories in the $s$-ensemble}
\label{sec:traj}

We have proven the existence of a first-order dynamical
phase transition in KCMs, and compared the behaviour of these
models with the A and AA models.  The effect of the field $s$
is to generate an ensemble of histories, biased towards
small or large activity.  In order
to gain insight into this transition, we now discuss 
the histories that dominate the $s$-ensemble when $s$ is finite. 

\subsection{Effect of temporal boundary conditions in the $s$-ensemble}
\label{sec:tfin_vs_tint}

\subsubsection{General considerations}

In steady states, the (unbiased) ensemble of histories
is invariant under translation in time.  Suppose that
$b=b(\C)$ is a configuration-dependent observable, and 
$B=\int_0^t\dd t'\, b(\C(t'))$.  Then, for trajectories of length $t$ 
the expectation value of the observable $b$ at time $\tau$ is
\begin{equation}
\langle b(\tau) \rangle = \frac 1t \langle  B\rangle  ,
\end{equation}
independent of the time $\tau$.  

However, introducing a field $s$ biases the ensemble
of histories, and, in general, time translation invariance
is broken.  This
effect is a dynamical analog of boundary effects in
classical thermodynamics:  if a system is finite,
the behaviour near its boundaries is
different from that of the bulk.  In the $s$-ensemble,
we consider trajectories $\C(\tau)$: the boundaries
of the trajectory are $\tau=0$ and $\tau=t$, while
the analogy of the `bulk' is $0\ll\tau\ll t$.  In the limit
of large time, extensive quantities are dominated by the
bulk: we have
\begin{equation}
\langle b(\tau) \rangle_s = \frac 1t \langle  B\rangle_s,
\end{equation}
for $0\ll\tau\ll t$.  However, in general we have
$\langle b(\tau) \rangle_s\neq\langle b(t) \rangle_s
\neq\langle b(0) \rangle_s$.
(In section~\ref{sec:modelA_rho_s_tau} we illustrate these
differences by calculating $\langle b(\tau)\rangle$ in the A-model.)

More generally, it is possible to express the average at the
final time,
$\langle b(t) \rangle_s$ and the time average $\frac 1t \langle B
\rangle_s$ by considering the eigenvalues and
the eigenvectors of the operator $\mathbb W_A$
(see appendix~\ref{app:eigenvecW_A}).
Using this approach, one can perform a perturbation theory around
$s=0$. In particular, when detailed balance is verified
and $s$ is conjugate to an observable of type $B$,
the bulk and boundary averages differ at first order in $s$:
for large times
\begin{eqnarray} \label{eq:B_and_b}
    \langle b(t) \rangle_{s} &=&   \langle b \rangle + 
  s b^{(1)} + \dots 
\\
    \langle B \rangle_{s} &=&  t\left[ \langle b \rangle + 2s b^{(1)} + \dots \right],
\end{eqnarray}
where an explicit expression for $b^{(1)}$ is given in Eq.~\eqref{eq:b1}.

\subsubsection{Effects of temporal boundaries in the A-model}
\label{sec:modelA_rho_s_tau}

We now illustrate the effect of temporal boundaries
in the $s$-ensemble, using the (bosonic) A-model.
We define the average particle density 
in this ensemble at a time $\tau$: that is, 
\begin{equation}
\rho(s;\tau,t)=\frac 1\vol \langle n_\mathrm{tot}(\tau)\rangle_s,
\end{equation} with $0\leq\tau\leq t$
and $n_\mathrm{tot}=\sum_i n_i$.  
To completely specify the problem, we must 
set the initial conditions in~\eqref{eq:hist_measure}: we take
a Poisson distribution with mean density $c_0$:
\begin{equation} \label{eq:a_mod_ic}
p_0(\C_0) = \prod_i \frac{c_0^{n_i} \ee^{-c_0}}{n_i!}
\end{equation}

To proceed, we write
$\mathcal{N}_\tau[\mathrm{hist}]=\int_0^{\mathrm{min}(\tau,t)}\dd t' 
  n_\mathrm{tot}(t')$, and we define
$P(n_\mathrm{tot},\mathcal{N}_\tau,K,t)$ to be the probability
that the system contains $n_\mathrm{tot}$ excitations
at time $t$, having made $K$ changes of configuration, and
with the observable $\mathcal{N}_\tau[\mathrm{hist}]$ 
taking a value $\mathcal{N}_\tau$.  Then, we define the 
generating function
\begin{equation}
\hat P_{n_\mathrm{tot}} \equiv \hat P(n_\mathrm{tot},h,s,t)=
    \sum_K \int \dd \mathcal{N}_\tau \ee^{-h \mathcal{N}_\tau -sK} 
       \hat P(n,\mathcal{N}_\tau,K,t)
\end{equation}
so that 
\begin{equation} \label{eq:rho_tau_gen}
  \rho(s;\tau,t) = -\frac 1\vol \frac{\partial}{\partial \tau} 
  \frac{\partial}{\partial h} \ln
  \sum_{n_\mathrm{tot}} \hat P(n_\mathrm{tot},h,s,t)
\Big|_{h=0} 
\end{equation}

Deriving an equation of motion for $\hat P_{n_\mathrm{tot}}$
is a straightforward generalisation of the derivation of~\eqref{eqn:evol_hatP}:
the result is
\begin{equation}
  \partial_t \hat P_{n_\mathrm{tot}} = 
  \ee^{-s}\big[c \vol \hat P_{n_\mathrm{tot}-1}+
   (n+1)\hat P_{n_\mathrm{tot}+1}\big] - 
  \big[c\vol +n+hn\Theta(\tau-t)\big] \hat P_{n_\mathrm{tot}}
\end{equation}
where $\Theta(t)$ is the usual Heaviside step function.  
With the initial condition of~\eqref{eq:a_mod_ic}, this
equation of motion is solved by a Poisson
distribution with a time-dependent normalisation factor: we take
$P(n_\mathrm{tot},h,s,t)=\exp[\Psi(t)-\rho_0(t)] 
\rho_0(t)^{n_\mathrm{tot}}/(n_\mathrm{tot}!)$.  Then, the mean, $\rho_0(t)$,
and normalisation factor, $\Psi(t)$, obey
\begin{eqnarray}
      \dot\rho_0(t)&=&c\vol \ee^{-s}-\big[1+h\Theta(\tau-t)\big]\rho_0(t) \\
      \dot\Psi(t)&=&\dot\rho_0(t)+\ee^{-s}\rho_0(t)-c\vol 
\end{eqnarray}
with initial conditions $ \rho_0(0)=c_0\vol $, $\Psi(0)=0$.
We identify $e^{\Psi(t)}=\sum_{n_\mathrm{tot}}
P(n_\mathrm{tot},h,s,t)$, so we solve for $\Psi(t)$ and 
use~\eqref{eq:rho_tau_gen} to obtain
\begin{equation}
  \rho(s;\tau,t)  =
    cz^2 + \ee^{-t} (1-z) (c_0-cz) + \ee^{-\tau} z (c_0-cz) + \ee^{\tau-t} cz(1-z)
\end{equation}

where we defined $z=\ee^{-s}$ for ease of writing.

\begin{figure}[t]
  \centering
  \includegraphics[width=0.49\columnwidth]{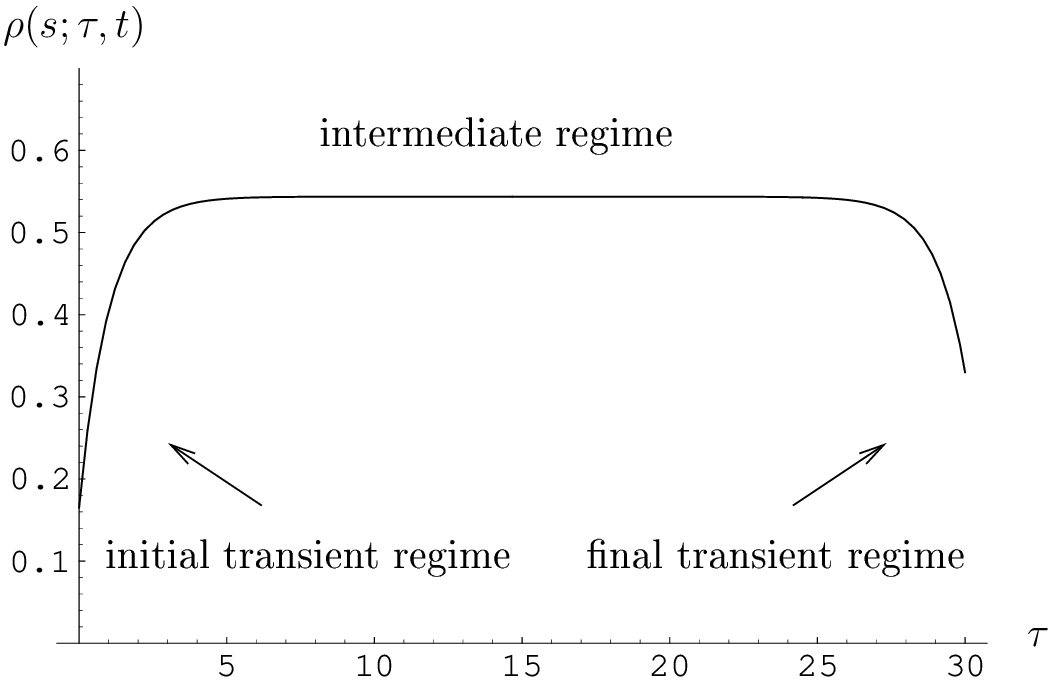}
  \caption{Mean density as a function of time in the A-model,
for the $s$-ensemble of histories of length $t=30$. 
We take $c_0=0.1$, $c=0.2$, $s=-0.5$.}
  \label{fig:transient_rho_tau_t}
\end{figure}

The $\tau$-dependent density $\rho(s;\tau,t)$ exhibits four
different regimes (Fig.~\ref{fig:transient_rho_tau_t}).
\begin{itemize}
\item Short trajectories. For $\tau<t\ll1$,
the system has a density $\rho(s;\tau,t)\simeq c_0$ close to 
the density at time $0$.

\item Long trajectories, stationary (bulk) regime.
For $1\ll \tau \ll t $,  the system adopts a density 
$\rho(s;\tau,t)\simeq c \ee^{-2s}$, independent of $\tau$.
For long trajectories, this average value coincides
with the time averaged density $t^{-1}\int_0^t \dd\tau \rho(s;\tau,t)$.

\item Long trajectories, initial transient regime.  For early times
$\tau\ll1\ll t$, the density depends on the value of $s$.  This dependence
persists even for $\tau=0$: that is,
the trajectories that dominate the $s$-ensemble
have non-typical initial conditions as well as non-typical
bulk properties.  To be precise, $\rho(s;0,t)=c_0\ee^{-s}$: the influence
of the initial condition decays into the bulk as 
$\rho(s;\tau,t)\simeq (c_0\ee^{-s}-c\ee^{-2s})\ee^{-\tau}+c\ee^{-2s}$.

\item Long trajectories, final transient regime: for $\tau\to t$ 
the density at the final time $t$ is
$\rho(s;t,t)=c \ee^{-s}$.  Moving away from this boundary,
the density decays into the bulk as
$\rho(s;\tau,t)\simeq (c\ee^{-s}-c\ee^{-2s})\ee^{\tau-t}+c\ee^{-2s}$.
\end{itemize}

We note that if the initial density $c_0$ is equal to the equilibrium
density $c$, then the ensemble at $s=0$ has time-reversal
symmetry.  Since the observable $K$ respects this symmetry,
the $s$-ensemble is also time-reversal symmetric $\rho(s;\tau,t)=
\rho(s;t-\tau,t)$, for $c=c_0$.  In this case,
the initial and final transient regimes are related
by this symmetry.

We have used the A-model to calculate the time dependence
of $\rho(\tau)$ exactly.  However, we emphasise that
the four regimes identified here are very general.  When
$t$ is large, histories in the $s$-ensemble are characterised by
an extended intermediate (bulk) regime, with initial and final
transient regimes that decay exponentially into the bulk.

\subsection{Landau-like theory for fluctuations within the $s$-ensemble}
\label{sec:DynLandauF}

In this section, we study large deviations of observables
within the $s$-ensemble.  For example, for an $s$-ensemble
parameterized by the observable $K$, we consider the probability
of observing a history with a particular value of an observable $B$ unrelated to $K$.  
In particular, we connect the large deviations
of the average excitation density $\rho$  to the variational
free energy $\mathcal{F}_K(\rho,s)$, defined in~\eqref{eq:def_F_rho}.  

\subsubsection{Variational calculation of $\psi_K(s)$ in a
general mean-field model}
\label{sec:var_psi_mf}

We consider systems for which
we can write the master-like equation~\eqref{eqn:evol_hatP}
in terms of a single co-ordinate $n_\mathrm{tot}$.  In
the mean-field FA model, this co-ordinate is the 
total number of excitations, but it might
also represent (for example), the total magnetisation
of a mean-field Ising model~\cite{FormaThermo}.
To be precise, we assume the master-like operator $\mathbb W_K$
has matrix elements
\begin{align} \label{equ:WK_bare}
  \big(\mathbb W_K\big)_{n,n'} \ = \ &
   \ee^{-s} W^{+}_{n-1}\,\delta_{n',n-1} 
   \ + 
   \ee^{-s} W^{-}_{n+1}\,\delta_{n',n+1} 
  - \big[W^{+}_n+W^{-}_n\big]  \delta_{n',n}.
\end{align}
where we have abbreviated $n_\mathrm{tot}$ to $n$, for compactness,
and $W^\pm_n$ are the rates for transitions
from the state $n$ to state $n\pm1$.  
For an example, consider the (bosonic) mean-field variant 
of the FA model, for which 
$W^+_n=W(n\to n+1)=cn$ and $W^-_n=W(n\to n-1)=(n-1)n/N$, as defined
in Section~\ref{sec:def_bos_FA}.  
We have assumed for convenience that
all processes in the system change the co-ordinate $n_\mathrm{tot}$
by one: the generalisation to other cases (such as the mean-field
AA model) is straightforward.

Following Section~\ref{sec:variational_approach}, we
now symmetrise the operator $\mathbb W_K$, so that
the dynamical free energy $\psi_K(s)$
is given by the largest eigenvalue of the operator
\begin{equation} \label{eq:WK_s}
(\tilde{\mathbb{W}}_K)_{n,n'} = 
  (W^+_n W^-_{n+1})^{1/2} {\rm e}^{-s} \delta_{n',n+1} +
  (W^+_{n'} W^-_{n'+1})^{1/2} {\rm e}^{-s} \delta_{n,n'+1} -
  (W_n^+ + W_n^-) \delta_{n,n'}
\end{equation}
For large systems, ($\vol \to\infty$) the eigenvector associated
with the largest eigenvalue takes the form 
$V_n=\ee^{-\vol f(\rho)/2}$ with $\rho=n_\mathrm{tot}/\vol $,
and the function $f(\rho)$ has a unique global minimum.
Then, Eq.~\eqref{eq:max_P} states that
\begin{equation} \label{eq:max_P_f}
  \psi_{K}(s) \ =\ \max_{f(\rho)} 
   \frac{\sum_{n_\mathrm{tot}} \Big\{ 
      \ee^{-s}[W^{+}_{n_\mathrm{tot}}W^{-}_{n_\mathrm{tot}}]^{1/2}\big(
        \ee^{-f'(\rho)}+\ee^{+f'(\rho)}\big)
                 -W^{+}_{n_\mathrm{tot}}-W^{-}_{n_\mathrm{tot}} 
                        \Big\} \,\ee^{-\vol f(\rho)}}
        {\sum_{n_\mathrm{tot}} \ee^{-\vol f(\rho)}}
\end{equation}
For any trial function $f(\rho)$, the sums over $n_\mathrm{tot}$ 
in~\eqref{eq:max_P_f} 
are dominated by the occupation numbers $n_\mathrm{tot}$ 
such that $f(\rho)$ is minimum (which implies in particular $f'(\rho)=0$). 
Thus, the direct dependence
on $f$ vanishes: we are left with a maximisation over the 
position of the minimum in $f(\rho)$.
Since the form of $f(\rho)$ is irrelevant, 
the eigenvector $V_n$ can be written in the form given
in Eq.~\eqref{eq:V_rho}.  Using this choice, we arrive at
\begin{equation}
\label{eq:minimalaLandau}
   \psi_{K}(s)  = -\min_\rho \mathcal F_K(\rho,s)
\end{equation}
where the variational free energy $\mathcal F_K(\rho,s)$
was originally defined in~\eqref{eq:def_F_rho}.  
For these mean-field models, it takes the form
\begin{equation}
\mathcal F_{K}(\rho,s) \ = \  
\frac 1 \vol  \{ -2 \ee^{-s}[W^{+}_{\vol \rho} W^{-}_{\vol \rho}]^{1/2} + W^{+}_{\vol \rho} 
+ W^{-}_{\vol \rho}\}
\end{equation}

As discussed in Section~\ref{sec:bosonic_FA_var}, 
$\mathcal F_{K}(\rho,s)$ gives a bound on
$\psi_K(s)$ for all systems.  However, 
for systems with mean-field geometry, we showed that the form
of the trial distribution is irrelevant in the limit of
large system size.  Thus, we write
Eq.~\eqref{eq:minimalaLandau} with an equality, and not as a bound.
We now discuss the physical interpretation of this result.

\subsubsection{Physical interpretation of the variational
free energy}

Consider an $s$-ensemble in which trajectories are
weighted by the usual factor
$\ee^{-sK[\mathrm{hist}]}$, but with the further restriction
that the time-integrated density be fixed.  
That is, we write
    the (unnormalized) probability, in the $s$-ensemble, to measure
    a time-averaged density $\rho$, 
\begin{equation}
\label{eq:defFstar}
 \Big\langle \ee^{-sK} \delta \Big[\rho-\frac{1}{\vol t}\int_0^t \dd\tau\: n(\tau) \Big]\Big\rangle 
\sim \ee^{-\vol t \mathcal F_{K}^\star(\rho,s)}
\end{equation}
where the asymptotic behaviour of the left hand side at large $t$
defines the function $\mathcal F_{K}^\star(\rho,s)$.  

Taking a Laplace transform of~\eqref{eq:defFstar} with respect to $\rho$,
we arrive at
\begin{equation} \label{eq:Fstar_generating}
Z_{K,\mathcal{N}}(s,h) \equiv
  \Big\langle \exp\Big(-sK-h\mathcal{N} \Big)\Big\rangle 
= \int \dd\rho\: \ee^{-(\mathcal F_{K}^\star(\rho,s)+h\vol t\rho)}
\end{equation}
where we write
$\mathcal{N}=\int_0^\tau\dd\tau\, n_\mathrm{tot}(\tau)$, noting
the similarities with
the generating function of~\eqref{eq:rho_tau_gen}.

Now, by analogy with~\eqref{eq:def_Z_A_Laplace}, we identify 
$Z_{K,\mathcal{N}}(s,h)$ as the partition function for
an `$(s,h)$-ensemble', in which histories are biased both by
their activity $K$ and their time-integrated number
of excitations $\mathcal{N}$.  Repeating the analysis of 
section~\ref{sec:s_ensemble_obsA_B}, we observe that
\begin{equation}
\psi_{K,\mathcal{N}}(s,h)=\lim_{t\to\infty} \frac 1t \ln
Z_{K,\mathcal{N}}(s,h)
\end{equation} 
is the largest eigenvalue of an operator $\mathbb{W}_{K,\mathcal{N}}$,
whose elements are
\begin{equation}
(\mathbb W_{K,\mathcal{N}})_{n',n} =
   W^{+}_{n-1}\,\ee^{-s}\delta_{n',n-1} 
      \ + 
   W^{-}_{n+1}\,\ee^{-s}\delta_{n',n+1} - \big[W^{+}_n+W^{-}_n+hn\big]  \delta_{n',n}
\end{equation}

The largest eigenvalue of this operator can be obtained by symmetrising
and repeating the variational analysis of the previous section.  
The result is
\begin{equation}
\psi_{K,\mathcal{N}}(s,h) =  - \min_\rho (\mathcal F_{K}(\rho,s)+h\rho) 
\end{equation}
which applies in the limit of large system size $N$ [since
in that case, the maximisation over the function $f(\rho)$
can be replaced by a maximisation over the density $\rho$].
However, performing a saddle point analysis directly
on~\eqref{eq:defFstar} reveals 
(for large times $t$ and
finite system size $N$),
\begin{equation}
\psi_{K,\mathcal{N}}(s,h) =  - \min_\rho (\mathcal F_{K}^\star(\rho,s)+h\rho) 
\end{equation}
Thus, in the limit of large system size $N$,
the Legendre transforms of $\mathcal F_{K}(\rho,s)$
and $\mathcal F_{K}^\star(\rho,s)$ are equal.  It follows that the
large deviation function $\mathcal F_{K}^\star(\rho,s)$
coincides with the variational free energy $\mathcal F_{K}(\rho,s)$
as long as the inverse Legendre transform can be performed,
However, in Section~\ref{sec:bosonic_FA_var}, 
we showed that in KCMs, $\mathcal{F}(\rho,s)$
typically has two minima, separated by a 
range of densities in which it is `non-convex':
$\partial_\rho^2 \mathcal{F}(\rho,s)>0$.
In this case, the inverse Legendre transformation cannot
be performed.  In fact, the non-convexity
of $\mathcal{F}(\rho,s)$ arises because histories with
some values of $\rho$ are unstable in the $s$-ensemble,
as we now show.

\subsubsection{Non-convex free energies: phase separation in time}

In the thermodynamics of finite-dimensional systems, one
typically has $s''(e)\leq0$\cite{Kuboetal}.  
Loosely, this property
arises because any energy density $e$ can be achieved by separating
a system into two regions, separated by an interface 
whose energy cost scales subextensively with the size of the
system.  Thus, the total energy density is $e=(1-x)e_1 + xe_2 + \delta$
where $e_1$ and $e_2$ are the energy densities of the two regions,
$x$ is the fraction of the volume of the system taken up
by the second region, and $\delta$ is the energy of the interface
divided by the total volume, which vanishes in the thermodynamic limit.
This leads to the `lever rule' $e=(1-x)e_1 + xe_2$.
The total entropy density associated with these configurations is
$s(e)=(1-x)s(e_1)+xs(e_2)$, and using the lever rule,
it follows that $s''(e)\leq0$.
However, in mean-field geometries, interfaces cannot be formed,
and this argument cannot be applied.

Interestingly, in the statistics of histories, phase separation
is possible even in mean-field systems.  We consider 
the large deviation function
$\mathcal F_K^\star(\rho,s)$, at a density $\rho$ for which
$\mathcal F_K(\rho,s)$
is non-convex.  We will find that the average
in~\eqref{eq:defFstar} is dominated 
by histories that phase separate \emph{in time}.
To prove this, we use the methods of Donsker-Varadhan theory,
described in appendix~\ref{app:DonskerVaradhan}.  
This method allows us to prove that
\begin{equation} \label{eq:var_Fstar}
\mathcal{F}_K^\star(\rho,s) = \left. \min_{|V\rangle} 
  \frac{\langle V|\tilde{\mathbb W}_K|V\rangle}{\langle V|V\rangle}
\right|_{\langle V|\hat\rho|V\rangle=\rho}
\end{equation}
where the minimisation is over distributions $V(\C)$ such that
$\sum_\C V(\C)^2 \rho(\C)=\rho$.  By analogy with
the thermodynamic case,
we take $V(\C)=(1-x)V_{\rho_1}(\C)+xV_{\rho_2}(\C)$, 
where $V_\rho(\C)$ was defined in~\eqref{eq:V_rho}.
We then minimise over the densities $\rho_1$ and $\rho_2$, 
choosing $x=(\rho-\rho_1)/(\rho_2-\rho_1)$
to ensure that the mean density is $\rho$.  Taking $x=0$,
we have a bound $\mathcal{F}_K^\star(\rho,s)
\leq \mathcal{F}_K(\rho,s)$.  However, if $\mathcal{F}_K(\rho,s)$
is non-convex (that is, $\partial_\rho^2 \mathcal{F}_K(\rho,s)<0$)
we can find a lower bound on $\mathcal{F}_K^\star(\rho,s)$
that is smaller than $\mathcal{F}_K(\rho,s)$.
For example, in the FA model in finite dimension (and
in the limit of large system size $N$), we find that
$\mathcal{F}^\star(\rho,s)$ is minimised by $\rho_1=0$,
$\rho_2=cz^2$ and $x=\rho/\rho_2$, for $0<\rho<c\ee^{-2s}$.
For $0<\rho<cz^2$, this variational approximation to 
$\mathcal{F}_K^\star(\rho,s)$ indicates that the system separates into
two phases with densities $0$ and $cz^2$.  We arrive at a
bound
\begin{equation}
\frac{\mathcal{F}_K^*(\rho,s)}{2d} \leq \left\{ \begin{array}{ll}
  \rho c(1-z^2), & \rho \leq cz^2 \\
  \rho (c+\rho-2z\sqrt{c\rho}), & \rho \geq cz^2 
  \end{array}\right.
\end{equation}
from which we note that $\partial_\rho^2\mathcal{F}^\star(\rho,s)=0$
in the two-phase regime: this is the Maxwell construction~\cite{Kuboetal}.
The result for the mean-field FA model is obtained by setting $2d=1$, as in
Section~\ref{sec:bosonic_FA_var}: in that case, the bound is saturated
(this follows since $\mathcal{F}^\star$ is convex
and its Legendre transformation is known to be equal to that of $\mathcal{F}$).

In addition to establishing the convexity of $\mathcal{F}_K^\star(\rho,s)$,
Donsker-Varadhan (DV) theory also provides an interpretation of the 
distribution $V^\star(\C)$ that minimises~\eqref{eq:var_Fstar}.
(We normalise to $\sum_\C V^\star(\C)^2=1$ for convenience.)
At large $t$, we consider 
histories in the $s$-ensemble with fixed average density $\rho$.
The DV theorem states that
this sub-ensemble is dominated by trajectories for which
the fraction of time spent in configuration $\C$ is 
$\mu^\star(\C)=V^\star(\C)^2$.
If the distribution
$\mu^\star(\C)$ is dominated by configurations $\C$ with density $\rho$, 
we conclude that the histories in this sub-ensemble are homogeneous in time.  
However, if $\mu^\star(\C)$ is associated with a bimodal density
distribution, it describes histories comprising separate
periods of of time, some with low excitation density $\rho_1$
and some with high excitation density $\rho_2$.  

Finally, we give the interpretation of 
$\mathcal{F}_K(\rho,s)$.  In systems
described by the single co-ordinate $n_\mathrm{tot}$, the DV theorem
states that $\ee^{-\vol t\mathcal{F}_K(\rho,s)}$ is 
the (unnormalised) probability (in the $s$-ensemble) of a history 
in which almost all configurations have density $(n_\mathrm{tot}/N)$
equal to $\rho$.  This
can be compared with the probability $\ee^{-\vol t\mathcal{F}_K^\star(\rho,s)}$
of a history with a time-averaged density $\rho$.  In this
sense, $\mathcal{F}_K(\rho,s)$ can be interpreted as a Landau-like
free energy for homogeneous trajectories, while $\mathcal{F}_K^\star(\rho,s)$
is the large deviation function for the density $\rho$.  In finite
dimension, $\mathcal{F}_K(\rho,s)$ provides a bound on the
large deviation function $\mathcal{F}_K^\star(\rho,s)$, based
on the assumption that histories are spatially and 
temporally homogeneous.  That is, fluctuations in space and
time are neglected.  In Section~\ref{sec:LandauFE_FT},
we discuss how these fluctuations can be taken into
account.

\subsubsection{Landau-like free energy in other $s$-ensembles}
\label{sec:dynLandauFE_MF}

So far, we have considered the large deviations of the density
$\rho$ in an $s$-ensemble that is defined in terms of the 
activity $K$.  The variational free energy can be simply
extended to $s$-ensembles defined as in~\eqref{eq:def_s_ens}.
Consider again a mean-field model specified by rates $W^\pm_n$
and an observable $A$ of the form given in~\eqref{eq:obsAsum},
which is incremented by $\alpha^\pm_n$ for transitions from state
$n$ to $n\pm1$.
Repeating the analysis of Section~\ref{sec:var_psi_mf},
we find $\psi_A(s)=-\vol\min_\rho \mathcal F_A(\rho,s)$ with
\begin{align} \label{eq:def_FA_m_s}
   \mathcal F_{A}(\rho,s) \ &= \  \frac 1 \vol  \big\{  
-2(W^{+}_{\vol \rho} 
  \ee^{-s(\alpha^+_{\vol \rho}+\alpha^-_{\vol \rho})} 
W^{-}_{\vol \rho}
)^{1/2} 
+ 
  W^+_{\vol \rho} + W^{-}_{\vol \rho}\big\}
\end{align}

For example, in the case of the complexity $Q_+$, 
one has $\alpha^\pm_n=\ln \frac{W^\pm_n}{W^+_n+W^-_n}$ and
\begin{equation} \label{eq:result_Qplus} \psi_+(s) =
    -\min_{\rho} \left\{-2\left[W^+_{\vol \rho} W^-_{\vol \rho} \right]^{\frac{1-s}{2}}
      \left[W^+_{\vol \rho} + W^-_{\vol \rho} \right]^{s} + W^+_{\vol \rho} + W^-_{\vol \rho} \right\} 
\end{equation}
where, again, this variational bound is exact because of its
independence of the form of the trivial wavefunction used.  By analogy
with $\mathcal{F}_K(\rho,s)$, we find that
$\ee^{-\vol t \mathcal{F}_A(\rho,s)}$
give the probability of homogeneous histories with density $\rho$,
in the $s$-ensemble.

\subsection{Dynamical free energy landscape (field theoretic approach)}
\label{sec:LandauFE_FT}

In Section~\ref{sec:DynLandauF}, we have considered
large deviations of time-averaged observables,
using these quantities to characterise the histories
within the $s$-ensemble.  We now discuss the calculation
of dynamical correlation functions within this ensemble.
We make use of a field-theoretic description of the FA
model.

\subsubsection{Field-theory for the bosonic FA model}

Using the Doi-Peliti representation of the bosonic
FA model (Section~\ref{sec:def_bos_FA}), we use
coherent states to write
the partition function $Z_K(s,t)$ as a path
integral over (time-dependent) functions $\{\hat\varphi_i\}$ and
$\{\varphi_i\}$~\cite{Doi-Peliti}.  Then, taking the continuum limit,
we promote these functions to fields $(\phi_{x\tau},\hat\phi_{xt})$ 
depending on position $x$ and time $\tau$, where $\phi_{x\tau}$
has the dimensions of a density and $\hat\phi_{x\tau}$
is dimensionless.  Introducing
sources $h$ and $\hat h$ for the fields $\phi$ and $\hat\phi$,
we write
\begin{equation}
\mathcal{Z}[s,t;h_{x\tau},\hat h_{x\tau}] 
  = \int\mathcal{D}[\phi_{x\tau},\hat\phi_{x\tau}] \exp\left\{
-S_K[\phi,\hat\phi]+\int\dd x
\dd\tau\, (h_{x\tau}\phi_{x\tau} + \hat h_{x\tau} \hat\phi_{x\tau}) \right\}
\end{equation}
where the path integral is over histories of duration $t$, and
(see, for example~\cite{Jack-Mayer})
\begin{equation}
S_K[\phi_{x\tau},\hat\phi_{x\tau}] = 
  \int\dd x \dd\tau\, \left\{ \hat\phi_{x\tau} \partial_t \phi_{x\tau}
 - 2d l_0^d \left[(\hat\phi_{x\tau}\phi_{x\tau}+cl_0^{-d}) - 
   \ee^{-s}(\phi_{x\tau}+cl_0^{-d}\hat\phi_{x\tau})\right] 
  (1+l_0^2\nabla^2)\hat\phi_{x\tau} \phi_{x\tau} \right\}
\end{equation}
where $l_0$ is the lattice spacing,
and we have taken a gradient expansion, truncating 
at quadratic order.
We identify $Z_K(s,t)=\mathcal{Z}[s,t;0,0]$.


\subsubsection{Saddle point approximation}
\label{sec:dynLandauFE_saddlepoint}

We now show that a saddle-point analysis on the action
recovers the results of the previous sections.
The saddle-point equations are obtained by maximising
the action with respect to $\phi$ and $\hat\phi$ in the
absence of the sources $(h,\hat h)$.  The
saddle occurs for fields that
are homogeneous in space and time, with magnitudes
satisfying
\begin{eqnarray}
0 &=& 2 \hat\phi\phi \left( \phi l_0^d - e^{-s} c \right) + \phi \left( c -
      e^{-s} \phi l_0^d \right) , \\ 0 &=& 2 \left( \hat\phi- e^{-s} \right)
      \hat\phi\phi + c \hat\phi\left( 1 - e^{-s} \hat\phi\right) .
\end{eqnarray}
These reduce to a single equation
if we take  $cl_0^{-d} \hat\phi=\phi$ (this origin of this
symmetry becomes clear if we use the symmetrised operator
$\tilde{\mathbb W}$ in the construction of the 
original path integral).  In this single variable, the
solutions are $\hat\phi=0$ and  
\begin{equation}
\hat\phi=\hat\phi_\mathrm{act} \equiv \frac{3}{4} e^{-s} + \frac{1}{4} \sqrt{9 e^{-2s} - 8}
\end{equation}
To estimate the dynamical free energy, we simply identify
$\psi_K(s)$ with $(-t^{-1} \min S[\phi,\hat\phi])$ where
the minimum is over value of the action
at the two saddles.  The result is
\begin{equation} \label{eq:psi_FT}
\psi_K(s) \simeq 
\left\{ \begin{array}{ll} 0, & s>0 \\
            \vol d (c\hat\phi_\mathrm{act})^2
       [\hat\phi_\mathrm{act}\ee^{-s}-1], & s<0 \end{array} \right.
\end{equation}
where the approximate equality indicates that we are working
in the saddle-point approximation.
We identify the time-dependent density (per site)
of excitations in the
$s$-ensemble
$\langle \rho(\tau)\rangle_s = l_0^{d} \langle \phi^*(\tau) \phi(\tau)\rangle$.
Away from temporal boundaries, we take the saddle point value
for this average, obtaining
\begin{equation} \label{eq:rho_FT}
\rho_K(s) \simeq 
\left\{ \begin{array}{ll} 0, & s>0 \\
            c(\hat\phi_\mathrm{act})^2, & s<0 \end{array} \right.
\end{equation}
It is easily verified that~\eqref{eq:psi_FT} and~\eqref{eq:rho_FT}
coincide with the variational estimates~\eqref{eq:psiK_FA_MF}
and~\eqref{eq:rhoK_FA_MF}. 

In principle, we can 
can now use the tools of dynamical
field theory~\cite{Zinn-Justin} to
incorporate fluctuations around the saddle
points, and to calculate spatiotemporal correlation
functions in the $s$-ensemble,
For example, defining 
a density field $n(x,\tau)$ through
a continuum limit of the original occupation variables $n_i$,
we have
\begin{equation}
\langle n(x,\tau) n(y,\tau') \rangle_s
 = \left. 
\frac{\delta^4}{\delta h(x,\tau) \delta h(y,\tau') \delta \hat h(x,\tau)
\delta \hat h(y,\tau')} \ln \mathcal{Z}[s,t;h,\hat h] \right|_{h=\hat h=0}
\end{equation}

Thus, for models with a field-theoretic
representation (such as the FA model),
the framework described in
this section provides methods for systematic calculation
of correlation functions and fluctuation effects in
the $s$-ensemble.  However, these
field-theoretical calculations beyond the scope of this paper.
We emphasize that the analysis of Sections~\ref{sec:phasecoexist}
and~\ref{sec:numericalmethod} establishes that
a dynamical first-order transition does occur at $s=0$
in finite-dimensional KCMs.  Thus, while we expect 
fluctuations to have quantitative effects, the qualitative
picture obtained through this saddle point analysis is
not changed.

\section{Outlook}
\label{sec:concl}

We have analyzed the dynamics of kinetically
constrained models, using an ensemble of histories
which span a long time $t$.
This analysis used dynamical
tools~\cite{Ruelle,Touchette,GaspardBook,Merolle-Jack,FormaThermo} constructed
by analogy with the usual Boltzmann-Gibbs theory of equilibrium
systems.
We have established that this procedure captures physically relevant
features that are not accessible from the steady state 
distribution of configurations in these models.

We have shown that the steady state of KCMs lies on a first-order
dynamical transition line, characterised by a coexistence between active and
inactive histories.   This first-order line
is present both in mean-field systems and in finite-dimensional
models.  Its existence is proven by variational bounds
on the dynamical free energy, and confirmed in numerical
simulations of several kinetically constrained models,
including both spin-facilitated models and kinetically constrained
lattice gases.  We have defined dynamical
Landau-like free energy, whose form is intimately connected 
to the existence of dynamical heterogeneities.

Earlier studies of non-equilibrium systems used a similar thermodynamic
formalism for dynamics to reveal first-order transitions arising from a
static phase transition~\cite{Ritort-04,Imparato-Peliti-05} or from an
absorbing state~\cite{FormaThermo}. To place our work in context, we
emphasise that our dynamical phase coexistence is not related
to such phenomena.  However, the transitions
in these models all appear as singularities in their large deviation
functions, consistent with the idea~\cite{Gaveau98} that phase
transitions both in and out of equilibrium can be studied through the
eigenvalue spectra of their master operators.  Moreover, the focus
of the current paper is on transitions between stationary, 
time-reversible dynamical states, and therefore we concentrated on
large deviations of quantities that are symmetric
in time: this is to be contrasted with studies that have 
concentrated on currents of entropy or 
particles~\cite{Kurchan-98,lebowitzspohn},
although recent work has hinted that large deviations of
time-symmetric observables may also be of importance in non-equilibrium
steady states~\cite{MaesTraffic}.

We expect our approach to be meaningful in a wider class of systems
than those probed in this paper.  For example, 
glass-forming liquids are known to be dynamically heterogeneous,
and this feature can be captured in computational
simulations of atomistic models. It would
be interesting to establish whether this heterogeneity is linked
to a dynamical phase transition similar to that present in KCMs.
This could
indicate a more general link between glassy properties (not
necessarily related to dynamical heterogeneity) and dynamical phase
transitions.

Finally, we observe that an experimental scheme for sampling the
$s$-ensemble would be very valuable, since it would provide a direct
test for the existence of a dynamical phase transition.  However,
the fact that the generalised master operator $\mathbb W_A$ does
not conserve probability makes the search for such a scheme
rather challenging.

\acknowledgments 
VL would like to thank Julien Tailleur for useful and continuous discussions, and 
RLJ and JPG thank David Chandler for extensive discussions.
This work (EP and FvW) was supported
by the French Ministry of Education through Grant No ANR-05-JCJC-44482.  JPG was supported by 
EPSRC under Grant No. GR/S54074/01.  While at Berkeley, RLJ was 
supported initially by NSF grant CHE-0543158 and later
by Office of Naval Research Grant No. N00014-07-1-068.  
VL was supported in part by by the Swiss FNS, under MaNEP
and division II.

\appendix

\section{Averages in the $s$-ensemble and eigenvectors of $\mathbb W_A$}
\label{app:eigenvecW_A}

\subsection{Eigenvectors of $\mathbb{W}_A$}

In this appendix, we discuss some properties of the
operator $\mathbb W_A$, and their consequences for averages
in the $s$-ensemble.
We write $\mathbb W_{A}$ in terms of its left
and right eigenvectors $|L_{n}\rangle$ and $|R_{n}\rangle$:
$
   \mathbb W_{A} = \sum_{n} \lambda_{n} |R_{n}\rangle \langle L_{n}|
$
with eigenvalues $\lambda_{0} > \lambda_{1} \geq \ldots $. 
The maximal eigenvalue
$\lambda_{0}$ is equal to $\psi_{A}(s)$.  
One can normalize eigenvectors so that
 \begin{equation} \label{eq:normLR}
 \langle L_{n}|R_{m}\rangle  = \delta_{nm}   \qquad\text{and}\qquad 
 \langle -|R_{0}\rangle  = 1
\end{equation}
where $\langle-|=\sum_{\C}\langle\C|$ is the projection state. 

Thus, for long times, we have
$\ee^{\mathbb{W}_At}=|R_0\rangle\langle L_0|e^{t\psi_A(s)} + \dots$
where the omitted terms on the right hand side
are exponentially smaller than the dominant first term.
Therefore,
starting from an initial state $|P_{0}\rangle=\sum_{\C}P_{0}(\C)|\C\rangle$,
with $\langle - | P_0\rangle=1$, one has, for large times
\begin{equation} \label{eq:P_long_t}
|P(t)\rangle=e^{\mathbb W_A t}|P_0\rangle \sim 
 |R_0\rangle \ee^{\psi_A(s) t} \langle L_0 | P_0 \rangle + \dots
\end{equation}
where we write the largest eigenvalue of $\mathbb W_A$
as $\lambda_0=\psi_A(s)$, and the omitted terms
are exponentially smaller than the first one,
for large times $t$.  This allows us to identify
the largest eigenvalue of $\mathbb W_A$ with the
dynamical free energy $\lim_{t\to\infty}t^{-1}\log Z_A(s,t)$,
through Equ.~\eqref{eq:Z_A_conv}.

\subsection{Time averages}

We now consider a configuration-dependent observable $b(\C)$,
and an $s$-ensemble defined as in~\eqref{eq:def_s_ens}, using an observable
$A$ of the form given in~\eqref{eq:obsAsum}.
We provide a link between the eigenvectors of $\mathbb W_A$ and 
two weighted averages:
  the average of $b$ at the final time $t$ in the $s$-ensemble 
  \begin{equation}
    \langle b(t) \rangle_{s} \equiv \frac{\langle b(\C(t)) \ee^{-sA}\rangle}{\langle \ee^{-sA}\rangle}
  \end{equation}
and the time-integrated average of $b$ in the $s$-ensemble
  \begin{equation}
    \langle B \rangle_{s} \equiv \frac{\left\langle \int_{0}^{t} d\tau\: b(\C(\tau)) \ee^{-sA}\right\rangle}
  {\langle \ee^{-sA}\rangle}
  \end{equation}

As discussed in Section~\ref{sec:tfin_vs_tint},
 $\langle B \rangle_{s} $ grows linearly in time, but, in general
$
   \partial_{t} \langle B \rangle_{s} \neq \langle b(t) \rangle_{s}
$.
In operator notation, the average of $b$ at the final time is
\begin{equation}
  \langle b(t) \rangle_{s} = \frac{\langle -|\hat b\, \ee^{t\mathbb W_{A}}|P_{0}\rangle}
  {\langle -|\ee^{t\mathbb W_{A}}|P_{0}\rangle}
\end{equation}
where $\hat b$ denotes the diagonal operator of elements $b(\C)$.
Using the normalization~\eqref{eq:normLR}, and the
large time result~\eqref{eq:P_long_t}, we arrive at
\begin{equation} \label{eq:moy_b_tfin}
  \langle b(t) \rangle_{s} = \langle -| \hat b|R_{0}\rangle
\end{equation}
Thus the right eigenvector $|R_{0}\rangle$ gives the distribution
over configurations $\C$ at the final time $t$.

On the other hand,
the integrated average $\langle B \rangle_{s}$ is obtained from the mean value
$
  \langle b(\tau) \rangle_{s}  
$
in the intermediate regime $0 \ll \tau \ll t$:
\begin{equation}
  \frac 1 t \langle B \rangle_{s} = 
  \langle b(\tau) \rangle_{s} =
  \frac{\langle -| \ee^{(t-\tau)\mathbb W_{A}} \hat b\, \ee^{\tau\mathbb W_{A}}|P_{0}\rangle}
  {\langle -|\ee^{t\mathbb W_{A}}|P_{0}\rangle}
\end{equation}
For $0 \ll \tau \ll t$, we have
\begin{align}
  \ee^{\tau\mathbb W_{A}}|P_{0}\rangle &\ = \ \ee^{\tau\psi_{A}(s)} |R_{0}\rangle\:\langle L_{0}|P_{0}\rangle
\\
  \langle -| \ee^{(t-\tau)\mathbb W_{A}} &\ = \ \langle - |R_{0}\rangle \: \langle L_{0}| \ee^{(t-\tau)\psi_{A}(s)}
\end{align}
and hence
\begin{equation} \label{eq:moy_b_tinterm}
  \frac 1 t \langle B \rangle_{s} = 
  \langle b(\tau) \rangle_{s} =
  \langle L_{0} |\hat b|R_{0} \rangle
\end{equation}
Thus, while the average $\langle b(t)\rangle_s$
depends only on $|R_0\rangle$, 
the average $\langle B \rangle_s$ depends on both $|R_0\rangle$
and $\langle L_0|$.

\subsection{Dynamics with detailed balance}

From~\eqref{eq:symmOp}, it follows that if a system obeys detailed balance,
its master operator satisfies
$\mathbb W_K^{\dag} = \hat P_{\text{eq}}^{-1}\mathbb W_K \hat P_{\text{eq}}$,
where $\hat P_\mathrm{eq}$ is a diagonal operator with elements $P_\mathrm{eq}(\C)$.
Thus,
$
  |L_{n}\rangle  =  \hat P_{\text{eq}}^{-1}|R_{n}\rangle
$.
Using this property together with results from the previous
section, and denoting 
$| R_0\rangle = \sum_{\C}R_0(\C,s) |\C\rangle$
we write
\begin{eqnarray}
    \langle b(t) \rangle_{s} &
=& \sum_{\C}b(\C) R_0(\C,s)
\\
    \partial_{t} \langle B \rangle_{s} &=  &
     \sum_{\C}b(\C) \frac{R_0(\C,s)^{2}}{P_{\text{eq}}(\C)}
\end{eqnarray}
Clearly, these averages are not the same in general.
Expanding about $s=0$ 
and using $R_0(\C,0)=P_\mathrm{eq}(\C)$,
we arrive at~\eqref{eq:B_and_b}, 
with
\begin{equation} \label{eq:b1}
b^{(1)} = 
     \sum_{\C}b(\C) \frac{\partial R_0(\C,s)}{\partial s}
\end{equation}

Finally, we note that expectation values of the form
$\partial_{t} \langle B \rangle_{s}$ take a simple form
when written in terms of the eigenvectors $|V\rangle$
of the symmetric operator $\tilde{\mathbb W}_K$, discussed
in Section~\ref{sec:variational_approach}.  The 
matrix elements of this operator are
$(\tilde{\mathbb W}_K)_{\C,\C'} = 
  P_{\text{eq}}^{-1/2}(\C) (\mathbb  W_K)_{\C,\C'} P_{\text{eq}}^{1/2}(\C)$,
so its eigenvectors are $V_n(\C,s)\propto \sqrt{L_n(\C,s)R_n(\C,s)}
=R_n(\C,s)/\sqrt{P_\mathrm{eq}(\C)}$.  Thus, we have (for large time)
\begin{equation} \label{eq:ave_B_V}
\frac 1t \langle B \rangle_{s} =
 \frac{ \sum_\C b(\C) V_0(\C,s)^2 }{ \sum_\C V_0(\C,s)^2}
\end{equation}
which links the eigenvector $V_0(\C,s)$
to physical observables such as $B$.

\section{Observables of types A and B}
\label{app:AB}

Here, we discuss the connections between observables of the 
forms given in~\eqref{eq:obsAsum} and~\eqref{eq:obsBsum}:
we refer to these observables as type A and type B respectively.
We begin with a result that is used in the numerical
methods of~\cite{Giardina,clones}.

Consider an $s$-ensemble defined as in Section~\ref{sec:s_ensemble_obsA_B}.
That is, take a system with rates $W(\C\to\C')$ and modify
the statistical weights of its histories by a factor $e^{-sA}$, where $A$ is 
an observable of type A.
In addition, we define a second stochastic process (`modified dynamics')
through the transition rates
\begin{equation} \label{eq:rates_Ws}
W_s(\C\to\C') = e^{-s\alpha(\C,\C')} W(\C\to\C')
\end{equation} 
where the $\alpha(\C,\C')$ are obtained from the definition
of the observable $A$, through~\eqref{eq:obsAsum}.
In addition we define two configuration-dependent
observables,
$r_s(\C) = \sum_{\C'} W_s(\C\to\C')$, and
\begin{equation}
\delta r_s(\C) = r_s(\C) - r(\C)
\end{equation}
[Here, $r(\C)=\sum_{\C'} W(\C\to\C')$ is the escape
rate for the dynamics $W$, as in the main text.] 

Motivated by the decomposition of
Equ.~\eqref{eq:decomp_dPdt_clones}, we can
establish two ways of defining the same
$s$-ensemble.  From~\eqref{eq:hist_measure}, it is easily
verified that
\begin{equation} \label{eq:equiv_ens}
\mathrm{Prob}[\mathrm{hist}|W] e^{-sA} = 
\mathrm{Prob}[\mathrm{hist}|W_s] e^{\delta R_s}  
\end{equation}
where the notation $\mathrm{Prob}[\mathrm{hist}|W]$ refers
to the (unmodified) probability of a history in a system with
dynamical rates $W$, and
\begin{equation}
\delta R_s = \int_0^t \dd\tau\, \delta r_s(\C(\tau))
\label{eq:delta_Rs}
\end{equation}
Thus,
Equ.~\eqref{eq:equiv_ens} states that histories in 
the $s$-ensemble parameterized by $A$ for the original dynamics $W$
have the same weight as histories in an $s$-ensemble
parameterized by $\delta R_s$, for the modified dynamics $W_s$.

Since the two ensembles are identical, it follows
that all observables have the same averages: for example
\begin{equation} \label{eq:equiv_AB}
\langle \mathcal{O} e^{-sA} \rangle_{W} = \langle
\mathcal{O} e^{\delta R_s} \rangle_{W_s},
\end{equation} where the
subscript on the average refers to the dynamical
rules used for the sum over histories.  Further,
this result holds for histories of finite duration
$t$, as long as the same initial conditions are
used in both averages.

For the specific case where the observable $A$ is the activity $K$ 
then this relation takes a particularly simple form.  Following
Section~\ref{sec:s_ensemble_obsA_B} with $\alpha(\C,\C')=1$ for
all $\C$ and $\C'$, the master operator
associated with this $s$-ensemble has matrix elements
$\big(\mathbb{W}_K\big)_{\C,\C'}=
    \ee^{-s}\,W(\C'\to\C)-r(\C)\delta_{\C,\C'}$ 
From~\eqref{eq:rates_Ws},
we find $W_s(\C\to\C')=\ee^{-s}W(\C\to\C')$: that is,
the modification to the
dynamics simply involves a rescaling of time by a factor
$\ee^{-s}$. 
In addition, for $B=R$, we have $\delta r_s(\C)=sr(\C)$, so
we define an
$s$-ensemble associated with the observable 
$R[\mathrm{hist}]=\int_0^t \mathrm{d}\tau\, r(\C(\tau))$, which is of 
type B.  From the analysis of Section~\ref{sec:obs_B},
the master operator associated with this ensemble, 
$\mathbb{W}_R$, has matrix elements
\begin{equation}
 (\mathbb W_{R})_{\C,\C'} = W(\C'\to\C) - (1+s) r(\C)
\delta_{\C,\C'}
\end{equation}
from which we can see that
\begin{equation}
\mathbb W_{K}(s) = \ee^{-s}\mathbb W_{R}\left(\ee^{s}-1\right)
\end{equation}

This equation relates the dynamical free energies
of the $s$-ensembles for $K$ and $R$, based on the
same unbiased dynamics $W$.  The dynamical
free energies $\psi_K(s)$ and $\phi_R(s)$ are
given by the largest eigenvalues of $\mathbb W_K$
and $\mathbb W_R$: they satisfy
\begin{equation}
  {\psi_{K}(s) = \ee^{-s}\phi_{R}\big(\ee^{s}-1\big)}
\end{equation}
Hence, we can also relate the cumulants of the observables
$K$ and $R$.  For example,
\begin{align}
\label{eq:K-R-mean}
  \langle K \rangle &=  \langle R \rangle \\
  \langle K^{2} \rangle_{c} &=  \langle R^{2} \rangle_{c}- \langle R \rangle
\end{align}
This last equation provides an interpretation of the 
variance (second cumulant) of $K$, through
\begin{align}
  \frac 1t \langle K^{2} \rangle_{c} 
  &= -\langle r \rangle +
     \int_{0}^{t}\: \frac{2t'}{t}
     \Big[\langle r(t) r(t')\rangle - \langle r(t)\rangle\langle r(t')\rangle\Big] \dd t'
\label{eq:link_K2_r}
\end{align}
where the correlation function is evaluated at $s=0$.

\section{Link to Donsker-Varadhan theory}
\label{app:DonskerVaradhan}

As in the main text, we
consider a Markov process described by transition rates
$W(\C\to\C')$ between configurations $\{\C\}$.  For a
history $\C(\tau)$, we define the
\textit{experimental measure}
\begin{equation}
  \bar \mu(\C,t) = \int_0^t d\tau\: \delta_{\C,\C(\tau)}
\end{equation}
This history-dependent observable simply counts how much time was
spent in configuration $\C$ between $0$ and $t$. This is the central
object of Donsker-Varadhan~\cite{DonskerVaradhan} theory
(see also~\cite{Touchette,Maesetal}).  
For large times, the experimental measure approaches the steady
state distribution,
$ \lim_{t\to\infty} \frac 1 t \langle
\bar\mu(\C,t) \rangle = P_{\text{st}}(\C)$.

Donsker-Varadhan theory gives information on the large deviations of
the experimental measure $\bar \mu(\C,t)$ from the steady-state
distribution, in the long time limit. Therefore,
it is naturally connected to the
statistics of histories and to the dynamical ensemble approach 
discussed in this article.
For example,
consider an observable $b(\C)$ depending on the
configuration of the system. 
The experimental measure $\bar\mu(\C,t)$
determines the time-integrated value of the observable $b$
through
\begin{equation} \label{eq:BmuCt}
\int_0^t d\tau \: b(\C(\tau)) =
  \sum_\C b(\C) \bar\mu(\C,t)  \equiv t \, \langle b \rangle_{\bar\mu}
\end{equation}
which defines $\langle b \rangle_{\bar\mu}$: the average
of $b$ with respect to the experimental measure $\bar\mu$.
In this appendix, we establish 
links between the $s$-ensemble approach and the results of Donsker
and Varadhan.  In particular, we develop a variational method that
gives the large deviations of an observable $B$, in the $s$-ensemble
defined for an (unrelated) observable $A$.

\subsection{Donsker-Varadhan large deviation function}

The Donsker-Varadhan (DV) theorem~\cite{DonskerVaradhan} states that in the
long time limit
\begin{equation} \label{eq:DV_thm}
  \text{Prob} \big[ \bar\mu(\C,t)= t \mu(\C) \big] = \ee^{t\,J[\mu]}
\end{equation}
with
(see for instance~\cite{Touchette,Maesetal})
\begin{equation}
  J[\mu] = \inf_{\rho > 0} 
    \sum_{\C,\C'} \Big\{ W(\C\to\C') \frac{\rho(\C')}{\rho(\C)}\,\mu(\C) -  r(\C) \mu(\C) \delta_{\C'\C}
    \Big \}
\end{equation}
where the infimum has to be taken over normalized measures $\rho(\C)$, with
$\sum_\C \rho(\C)=1$.

If $\mathbb W$ obeys detailed balance with respect to
an equilibrium distribution $P_{\text{eq}}(\C)$, 
the infimum is obtained for $\rho(\C)=\sqrt{\mu(\C)/P_\mathrm{eq}(\C)}$
and the large deviation function reduces to
\begin{equation} \label{eq:Jeq_mu_explicit}
  J_{\text{eq}}[\mu] = 
    \sum_{\C,\C'} \Big\{ 
    \big[ W(\C\to\C')W(\C'\to\C)\big]^{1/2} [ \mu(\C) \mu(\C') ]^{1/2}   -  r(\C) \mu(\C) \delta_{\C'\C}
    \Big\}
\end{equation}
for normalised measures $\sum_\C \mu(\C)=1$.  Writing
\begin{equation}
\label{eq:DV_V_mu}
\mu(\C)=\frac{V(\C)^2}{\sum_{\C} V(\C)^2},
\end{equation}
we identify
\begin{equation} 
  J_{\text{eq}}[\mu] 
= \frac{\langle V | \tilde{\mathbb W} | V \rangle}
  { \langle V |  V \rangle}
\end{equation}
as the function to be maximised in~\eqref{eq:max_P},
for the case $s=0$.

\subsection{Dynamical Landau free energy at $s=0$}

We now apply the DV theorem to the large deviations of an
observable $b(\C)$.  Integrating over a time $t$, we define
the history-dependent quantity
\begin{equation}
  B(t) = \int_0^t d\tau\: b(\C(\tau)),
\end{equation}
As
discussed in Section~\ref{sec:large_dev},
one expects the probability distribution of $B(t)$ to behave as
\begin{equation}
  \Omega_\mathrm{dyn}(B=bt,t) \sim \ee^{t\pi(b)}
\end{equation}
for large times $t$.

The large-deviation function $\pi(b)$ can be obtained through
the Donsker-Varadhan functional using 
$B(t) = \sum_\C b(\C) \bar \mu(\C,t)$, so that
\begin{align}
  \text{Prob} \big[ B(t) = t b \big] & = 
  \left \langle \delta\big(B(t) - t b \big)\right\rangle 
 \nonumber \\ &=
  \int d\mu \left \langle \delta\big(B(t) - t b \big)\:\delta\big(t^{-1} \bar \mu(\C,t) - \mu(\C) \big)\right\rangle 
 \nonumber \\ &=
  \int d\mu \: \delta\big(\text{$\sum_\C$} b(\C) \mu(\C) \,-\, b \big)\:
  \left \langle \delta \big(\bar \mu(\C,t) - t \mu(\C) \big)\right\rangle 
 \\ &=
  \int d\mu \: \delta\big( \langle b\rangle_\mu \,-\, b \big)\: \ee^{t J[\mu]} 
\end{align}
where the average $\langle b\rangle_\mu$ was defined in~\eqref{eq:BmuCt}.
Here, we have replaced an average over histories
$\langle\cdot\rangle$ with 
an integral over possible realisations of the experimental
measure $\mu$, weighted by their probabilities (which are known
from the DV theorem).  In the limit of large time, we 
maximise the argument of the exponential, subject to a
constraint imposed by the $\delta$-function.  Hence,
\begin{equation}
  \pi(b) = 
   \sup_{\substack{\mu ~\text{with} \\\langle b\rangle_\mu = b}} J[\mu]
\end{equation}
which for systems obeying detailed balance
can again be expressed in terms of the operator $\tilde{\mathbb W}$,
using~\eqref{eq:DV_V_mu}.

\subsection{Dynamical Landau free energy for any $s$}

We now generalise this analysis to the $s$-ensemble.  
We note that the values of `type A' observables [those
of the form given in~\eqref{eq:obsAsum}]
cannot be obtained from the experimental measure $\bar\mu(\C)$.  
To connect these observables to the DV approach, 
we use the results of appendix~\ref{app:AB}.  

The large deviations of the observable $B$ in the 
$s$-ensemble specified by $A$ are determined by 
\begin{equation}\Omega_A(s,b) = \langle \delta(B-bt) \ee^{-sA} \rangle_{W},
\end{equation}
where as in appendix~\ref{app:AB}, the label on the average
indicates the dynamical rules used to generate the ensemble
of histories.  From~\eqref{eq:equiv_AB}, we can write
\begin{equation}
\Omega_A(s,b) = 
\langle \delta(B-bt) \ee^{-sA} \rangle_{W} = \langle \delta(B-bt) 
  \ee^{\delta R_s} \rangle_{W_s},
\end{equation}
with an observable $\delta R_s$ and rates $W_s(\C\to\C')$ given
in Equ.~\eqref{eq:rates_Ws} and~\eqref{eq:delta_Rs}.

Now, following the analysis of the previous section, we have
\begin{align}
  \label{eq:Bweight_DV}
\Omega_A(s,b) =
  \Big\langle \ee^{\delta R_s} \delta \big(\bar B(t)-tB \big)\Big\rangle_{W_s}
   &=
  \int d\mu \: \delta\big( \langle b \rangle_\mu \,-\, B \big)\: \ee^{t \big(J[\mu|W_s]+\langle \delta r_s \rangle_{\mu}\big)} 
\end{align}
where $\langle \delta_r \rangle_\mu=\sum_\C \mu(\C) \delta r(\C)$, and
$J[\mu|W_s]$ is the Donsker-Varadhan functional for 
stochastic process with rates $W_s$.
Again, the integral over $\mu$ can be evaluated by maximising the argument
of the exponential subject to the constraint on $\langle b \rangle_\mu$,
leading to $\Omega_A(s,b) \sim \exp{t \pi_A(s,b) = } $ with
\begin{equation} \label{eq:LandauFE_nonMF}
 \pi_A(b,s) = - \sup_{\substack{\mu ~\text{with} \\\langle b\rangle_\mu = b}}  J_A[\mu,s] 
\end{equation}
with
\begin{equation} \label{eq:Jeq_A_mu_explicit}
  J_{A}[\mu,s] = 
    \sum_{\C,\C'}  \Big\{
     \ee^{-\frac s2 [\alpha(\C,\C')+\alpha(\C',\C)]}
     \big[W(\C\to\C')W(\C'\to\C)\big]^{1/2} [ \mu(\C) \mu(\C') ]^{1/2} 
             - r(\C) \mu(\C) \delta_{\C'\C}
     \Big\}
\end{equation}
where we emphasise that the rates $W(\C\to\C')$ are those
of the original (unmodified) dynamics.
From~\eqref{eq:DV_V_mu}, we identify
\begin{equation} \label{eq:expr_JAmus}
J_A[\mu,s] = \frac{\langle V|\tilde{\mathbb W}_A|V\rangle}{\langle V|V\rangle}
\end{equation}
as the quantity to be maximised in~\eqref{eq:max_P} for $A=K$. Moreover, for $b$ being the
occupation number $n$ and $A$ the activity $K$, one recognizes in~\eqref{eq:expr_JAmus}
the  result~\eqref{eq:var_Fstar}.

We observe that
these results have been derived for
dynamics which obey detailed balance, but they are not restricted to
that situation. For instance,~\eqref{eq:LandauFE_nonMF} holds in
general, with
\begin{equation} \label{eq:J_A_mu_explicit}
  J_{A}[\mu,s] = \inf_{\rho>0}
    \sum_{\C,\C'} \Big\{
     \ee^{- s\alpha(\C,\C')}
     W(\C\to\C') \frac{\rho(\C')}{\rho(\C)} \mu(\C)    -  r(\C) \mu(\C) \delta_{\C'\C}
     \Big\}
\end{equation}
Finally, we note that these results amount to a generalization of
the Donsker-Varadhan theorem~\eqref{eq:DV_thm} in the $s$-ensemble: for
large times,
\begin{equation} 
  \Big\langle \ee^{-sA} \delta\big( \bar\mu(\C,t)- t \mu(\C) \big) \Big\rangle= \ee^{t\,J_A[\mu,s]}
\end{equation}
with $J_A[\mu,s]$ given in general by~\eqref{eq:J_A_mu_explicit},
which reduces to~\eqref{eq:Jeq_A_mu_explicit} if $\mathbb W_A$ can
be symmetrised.

\end{document}